\documentclass[11pt]{article}
\usepackage{hyperref}
\usepackage{amsmath, amsfonts, amssymb, mathrsfs, dsfont}
\usepackage{dcolumn}
\usepackage{caption}
\usepackage{subcaption}
\usepackage{filemod}
\usepackage{natbib}
\usepackage[ruled, vlined]{algorithm2e}
\usepackage{floatrow}
\usepackage{setspace}
\usepackage{verbatim}
\usepackage{graphicx}
\usepackage{bbm} % for indicator function
\usepackage{xcolor} % only for coloring text while revising

\hypersetup{
    colorlinks=true,
    linkcolor=blue, 
    urlcolor=black,
    citecolor=blue, 
    }

\title{Bayesian Deep Gaussian Processes for Correlated Functional Data: \\
        A Case Study in Cosmological Matter Power Spectra}
\author{Stephen A. Walsh\thanks{Corresponding author: Division of Natural Sciences, 
        Math and Technology, Elms College, {\tt walshst@elms.edu}} \and 
        Annie S. Booth\thanks{Department of Statistics, Virginia Tech} \and
        David Higdon\footnotemark[2] \and
        Jared Clark\footnotemark[2] \and
        Kelly R. Moran\thanks{Los Alamos National Laboratory} \and
        Katrin Heitmann\thanks{Argonne National Laboratory}}
\date{\today}

\begin{document}

\maketitle
\bigskip

\begin{abstract} 
Understanding the structure of our universe and the distribution of matter is an 
area of active research.  As cosmological surveys grow in complexity, the development 
of emulators to efficiently and effectively predict matter power spectra is essential.  
We are particularly motivated by the Mira-Titan Universe simulation
suite that, for a specified cosmological parameterization (termed a ``cosmology''), 
provides multiple response curves of various fidelities, including correlated 
functional realizations.  Our objective is two-fold.  First, we estimate 
the underlying matter power spectra, with appropriate uncertainty 
quantification (UQ), from all of the provided curves.  To this end, we propose a 
novel Bayesian deep Gaussian process (DGP) hierarchical model which synthesizes 
all the simulation information to estimate the underlying matter power spectra
while providing effective UQ.  Our model extends previous work on Bayesian DGPs 
from scalar responses to correlated functional outputs.  Second, we leverage our predicted 
power spectra from various cosmologies in order to accurately predict the entire 
matter power spectra for an 
unobserved cosmology.  For this task, we use basis function representations 
of the functional spectra to train a separate Gaussian process emulator.  
Our method performs well in synthetic exercises and against the benchmark cosmological 
emulator (Cosmic Emu).
\end{abstract}


\section{Introduction}
%%%%%%%%%%%%%%%%%%%%%%%%%%%%%%%%%%%%%%%%%%%%%%%%%%%%%%%%%%%%%%%%%%%%%%%%%%%%%%%

\nocite{frontiere2022farpoint} % for coming fig.
Computer simulation experiments are invaluable tools in the study of cosmology.
Experiments that simulate the expansion of the universe are growing
in prevalence and complexity \citep[e.g.,][]{lawrence2010coyote,derose2019aemulus,
nishimichi2019dark,angulo2021bacco,euclid2021euclid,moran2023mira}.  
\begin{figure}[h]
    \centering 
    \includegraphics[width=.7\linewidth]{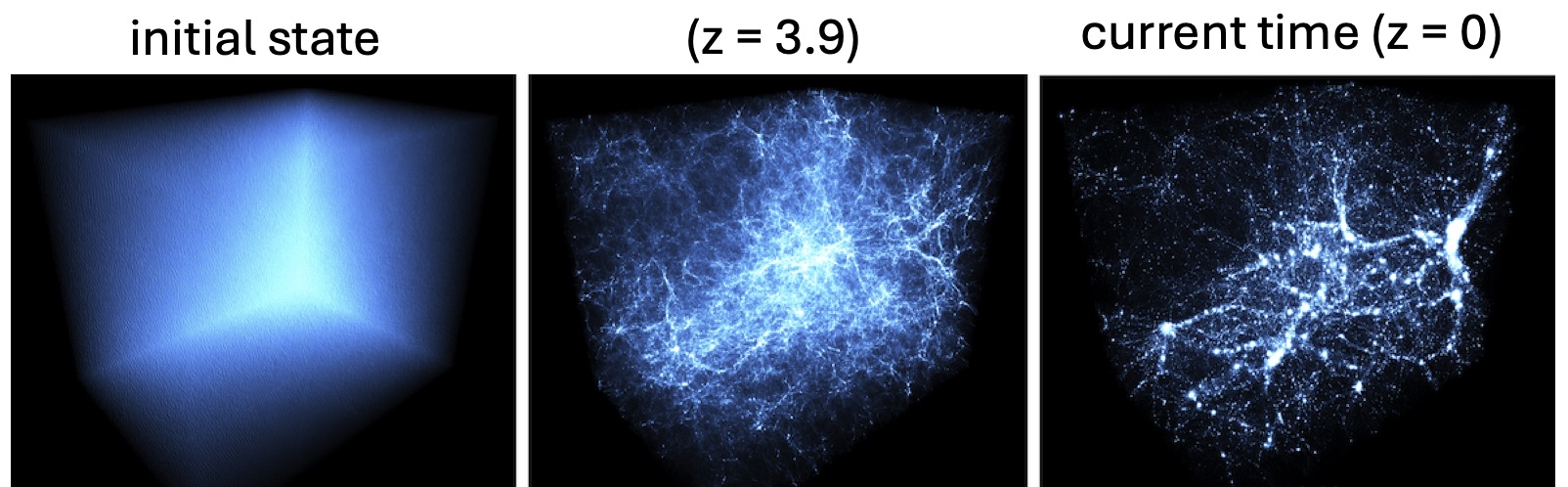}
    \caption{Snapshots of a large-scale structure simulation 
    evolving dark matter from shortly after the big-bang to now from \cite{frontiere2022farpoint}.  
    The simulations initialize and evolve over 30 billion matter particles according to gravitational forces.  
    The initialization is random, with density fluctuations consistent with early universe. 
    The simulation results in a matter density field over a periodic cube at the current time, 
    measured in redshift $z$, which decreases to $z=0$ for the present time.}
    \label{fig:matterSim}
\end{figure}
We are particularly motivated by simulations of the large scale structure of the 
universe (Figure \ref{fig:matterSim}). The simulations evolve dark matter particles 
from shortly after the big bang until now.  Different assumptions regarding cosmological 
parameters or the composition of the universe result in different spatial configurations 
of matter at the current time (redshift $z=0$).  Hence, comparing results from large 
scale surveys (e.g., the Sloan Digital Sky Survey, the Legacy Survey of Space and Time) 
with such simulations can serve as a check of theoretical understanding and can provide 
information about uncertain cosmological parameters.  Such simulations are also integral 
for planning and preparation for the arrival of data from upcoming surveys 
\citep[e.g.,][]{korytov2019cosmodc2, euclid2021euclid}.

The full simulation output is far too large to practically use for analysis.  
It is common to reduce the simulation output to its power spectrum.  For the final ($z=0$) 
configuration, this can be done efficiently by (1) computing a centered matter density 
field over a very dense three-dimensional grid $\delta(\mathbf{s})$, 
(2) taking the fast Fourier transform (FFT) $\widehat{\delta}(\mathbf{k})$, 
(3) taking the squared modulus $P(\mathbf{k}) = || \widehat{\delta}(\mathbf{k}) ||^2$, 
and (4) averaging over $\mathbf{k}$'s of common length 
$P(k) = \operatorname{ave}\limits_{||\mathbf{k}|| = k}
  P(\mathbf{k})$:
\[
\delta(\mathbf{s}) 
\stackrel{FFT(\cdot)}{\longrightarrow}
\widehat{\delta}(\mathbf{k})
\stackrel{||\cdot||^2}{\longrightarrow}
P(\mathbf{k})
\stackrel{\operatorname{ave}\limits_{||\mathbf{k}|| = k}}{\longrightarrow}
P(k).
\]
The FFT re-expresses the density field in terms of Fourier coefficients corresponding 
to different trignometric waveforms indexed by $\mathbf{k}$:
$
\widehat{\delta}(\mathbf{k}) = 
\sum_{s_1,s_2,s_3} \delta(\mathbf{s}) 
e^{-2\pi i \mathbf{k} \cdot \mathbf{s}/\ell}
$; 
$(s_1,s_2,s_3)$ indexes the three-dimensional array elements distributed over the 
cube with edge lengths $\ell$.  Hence the elements of $\mathbf{k}$ have units of 
$1/\ell$. We use the term wavenumber to refer to $k$. Wavenumber has an inverse 
relationship to spatial distance; small wavenumbers correspond to large scales 
such as galaxy clusters, while large wavenumbers correspond to small scales such 
as individual galaxies. The reduction of the spatial distribution of matter to its 
power spectrum is detailed in \citet{heitmann2010coyote}.

The spatial distribution of matter density  $\delta(\mathbf{s})$ is the result of 
a stochastic process that depends on the random initialization -- again, see 
\citet{heitmann2010coyote} for details. Also, $P(k)$ is the Fourier transform of 
the empirical covariance function of $\delta(\mathbf{s})$ \citep{schabenberger2017statistical}. 
Hence $P(k)$ is random, with variance that depends on $k$, the box size $\ell$ of the 
simulation, and the particle density.  This is visible in Figure \ref{fig:plot_raw_vs_scrP} 
which shows spectra resulting from simulations with different initializations, box sizes, 
and numbers of particles. We can also conceive of the ``infinite volume'' spectrum 
$P_\infty(k)$ which is the fixed (non-random) spectrum taken to be the limit of 
spectra produced by successively larger boxes with successively denser particles.
Estimating this infinite volume power spectrum as a function of cosmological parameters 
is the main analysis goal of this paper.

On large spatial scales ($k < 0.04$), 
gravitational interactions are sufficiently linear, so the evolution of matter structure 
can be described analytically via linear perturbation theory
\citep{pietroni2008flowing, lesgourgues2009non}.  Thus, the perturbation theory-based 
spectrum $P_{p}(k)$ accurately reproduces $P_\infty(k)$ for $k<0.04$.  On smaller spatial 
scales (i.e., for larger $k$), the nonlinear gravitational dynamics cannot be adequately 
captured by perturbation theory.  In this setting, the particle-based simulations of 
\cite{heitmann2016mira} give substantially more accurate matter evolution under gravitational forces.
The accuracy of the resulting simulation-based spectra depends on $k$, the box size $\ell$ 
and the density of matter particles used in the simulation.

The Mira-Titan simulation suite \citep{moran2023mira}
defines ``cosmologies'' according to eight cosmological parameters (more on this
in Section \ref{sec:data}).  For each cosmology this suite includes a single high-resolution 
simulation (box size $\ell = 2.1$Gpc (gigaparsecs), $3200^3$ particles), with spectrum $P_h(k)$, 
sixteen low-resolution simulations (box size $\ell = 1.3$Gpc, $512^3$ particles) with spectra 
$P_{\ell_r}(k),\,r=1,\ldots,16$, and a perturbation theory-based spectrum $P_p(k)$. The resulting 
spectra for a single cosmology are shown in Figure \ref{fig:plot_raw_vs_scrP}.
The box size limits the spatial scale at which the simulated spectrum accurately reflects 
$P_\infty(k)$.  For both the high and low resolution spectra, we can safely consider $P_h(k)$ 
and $P_\ell(k)$ unbiased for $k > 0.04$ Mpc$^{-1}$  (inverse megaparsecs). However, for low $k$ 
there is still a substantial amount of randomness in these spectra due to the initialization.  
We use these additional random, but unbiased, low-resolution spectra to give us better 
information about the infinite resolution spectrum $P_\infty(k)$.

Within a given volume, if the particle density is sufficiently large, non-linear gravitational 
forces are accurately reproduced in these simulations.  As volume decreases, the number of 
particles available to make the force calculations diminishes.  Eventually, these inaccuracies 
in force calculations for small distances will bias the resulting spectrum for large $k$. 
For the low-resolution Mira-Titan spectra shown in Figure \ref{fig:plot_raw_vs_scrP}, one can 
see this bias in the low-resolution spectra as compared to the high-resolution spectrum which 
has sufficent particle density (force resolution) to remain accurate for the larger values of 
$k$ plotted.

There are several challenges in estimating the infinite volume spectrum at each cosmology. 
One must account for randomness and dependence in the simulation-based spectra. 
One should also ensure the reconstructed infinite volume spectrum is smooth at large and 
small values of $k$, but still allows flexibility for intermediate values of $k$ 
($-1.4 \leq \log_{10}(k) \leq -0.8)$. We also aim to provide effective uncertainty 
quantification (UQ) regarding the estimated infinite volume spectrum, while incorporating 
expert knowledge regarding the values of $k$ over which various spectra are reliable.

To this end, we propose a Bayesian hierarchical model that treats functional simulation 
outputs (spectra) as realizations of a Gaussian process (GP) centered on the underlying 
infinite volume spectrum, which itself is modeled as a realization of a deep Gaussian process 
\citep[DGP;][]{damianou2013deep}.  The GP accommodates correlated observations
through its covariance structure.  The depth of the DGP offers nonstationary flexibility, 
allowing a smooth realization over some range of $k$, and a more variable realization in 
another.  The Bayesian framework facilitates UQ and the incorporation of prior
knowledge.  While Bayesian DGPs have been previously deployed for computer experiments 
with scalar responses \citep[e.g.,][]{sauer2023active,sauer2023vecchia,ming2023deep}, 
they have yet to be extended to correlated functional outputs. 
To demonstrate the proficiency of our Bayesian hierarchical DGP, 
we benchmark its performance in two unique settings: with synthetic data 
mimicking the Mira-Titan data and with predictions of the linear power spectrum from 
the Code for Anisotropies in the Microwave Background \citep[CAMB;][]{lewis2011CAMB}.
The CAMB model exhibits similar behavior to Mira-Titan but crucially
provides an infinite volume power spectrum for each cosmology for assessment.

Our second objective is to leverage simulation data from a limited set of
cosmologies in order to predict matter power spectra for unobserved cosmologies.
Due to the computationally expensive nature of the Mira-Titan simulation suite
(one batch of simulations can take multiple weeks to run on a supercomputer),
it is infeasible to evaluate the simulation for every possible eight-dimensional
cosmological configuration that may be of interest.  Instead, we desire
a statistical ``emulator'' or ``surrogate model'' 
\citep{santner2003design,gramacy2020surrogates} that will provide quick and effective
predictions of matter power spectra for any cosmological parameterization.
\citet{moran2023mira} provide a state-of-the-art emulator for the Mira-Titan
simulation suite, which was trained on 111 simulations, and is termed 
the ``Cosmic Emu.''\footnote{\url{https://github.com/lanl/CosmicEmu}}
We leverage the same training data, in conjunction with our Bayesian hierarchical
DGP model and basis function representations, to train a GP surrogate 
on principal component weights in order to predict spectra for unobserved
cosmologies \citep{higdon2008computer, higdon2010estcosmo}. 
Our method compares favorably to Cosmic Emu on held-out cosmologies.

The remainder of the paper is organized as follows.  Section \ref{sec:data} 
introduces our motivating application, the Mira-Titan simulation suite.  
Section \ref{sec:hm_fit} describes our hierarchical Bayesian DGP and 
validates its performance on synthetic exercises and the CAMB simulations.  
Leveraging this trained model, Section \ref{sec:pred} details 
our procedure for predicting at unobserved cosmologies, benchmarking against
state-of-the-art competitors on the CAMB and Mira-Titan simulations. 
Section \ref{sec:disc} concludes with a discussion of our contributions and avenues for 
future research.  We provide data, reproducible code, and an {\sf R} package for our 
Bayesian DGP model in a public git repository.\footnote{\url{https://github.com/stevewalsh124/dgp.hm}}

\section{Mira-Titan}
\label{sec:data}

The Mira-Titan simulation dataset consists of simulated matter power spectra for 
117 unique cosmologies (111 for training, 6 for testing). 
Each of these cosmologies is specified by eight cosmological 
parameters: matter density, baryon density, amplitude of density fluctuations, 
dimensionless Hubble parameter, spectral index of scalar perturbations,
dark energy equation of state parameters, and neutrino density. For more details 
on these parameters and their effects, see \cite{dodelson2020modern, aghanim2020planck, heitmann2016mira}. 
Parameter ranges are specified in Table 2.1 of \cite{moran2023mira}.

\begin{figure}[!t]
    \centering 
    \includegraphics[width=.85\linewidth]{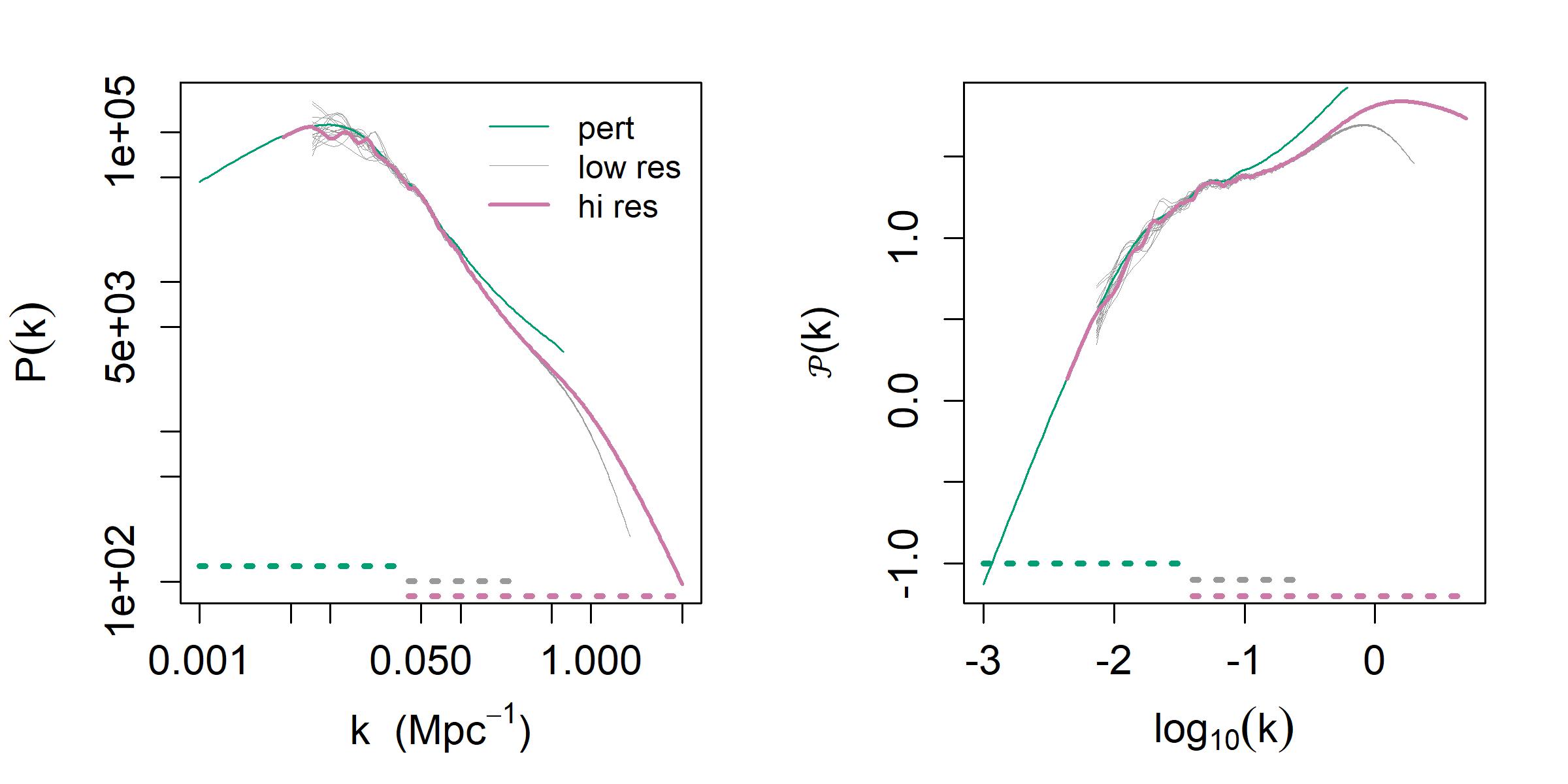}
    \caption{{\it Left:} The perturbation theory (solid green), low resolution runs (solid gray), 
    and high resolution run  (solid pink) on the original scale $P(k)$ for the first cosmology. 
    Dotted lines at the bottom
    indicate where each data type is deemed approximately unbiased. 
    {\it Right:} Same as left, but on the transformed space $\mathcal{P}(k)$ over $\log_{10}(k)$ values.}
    \label{fig:plot_raw_vs_scrP}
\end{figure}

\begin{figure}[ht]
    \centering 
    \includegraphics[width=.85\linewidth]{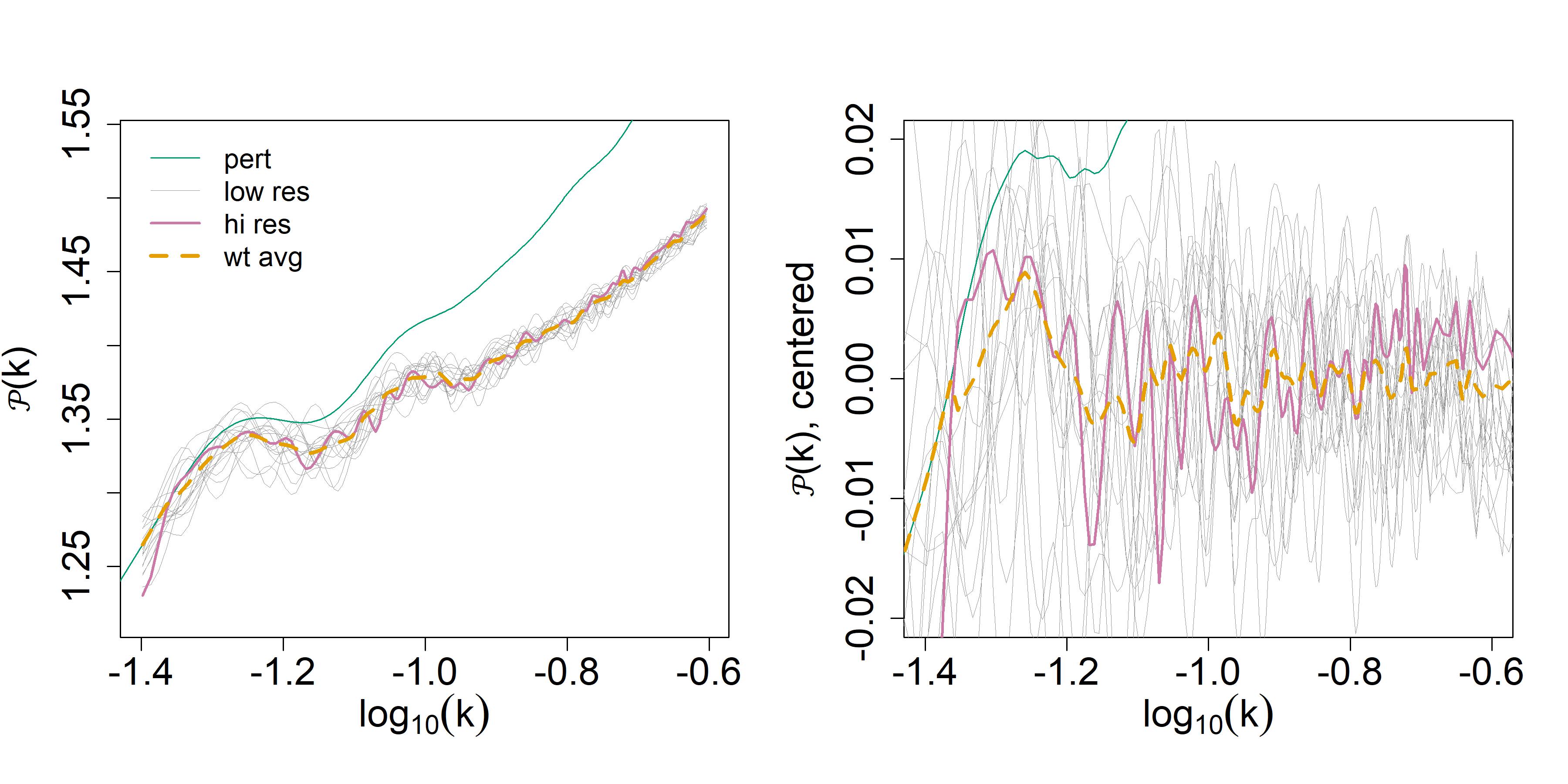}
    \caption{{\it Left:} The perturbation theory (solid green), low resolution runs (solid gray), 
    and high resolution run  (solid pink) for the first cosmology, restricted to wavenumber ($k$) 
    values where the low resolution runs are approximately unbiased, i.e., the span of the gray 
    dashed line in the left sub-figure.
    {\it Right:} Same as left, but with LOESS-smooth weighted average subtracted.}
    \label{fig:plot_data}
\end{figure}

Related approaches to multi-source cosmological emulation include GP-based 
multifidelity methods \citep{ho2022multifidelity, ho2023mfbox} and single-fidelity 
emulators such as PkANN II \citep{agarwal2014pkannII} and EuclidEmulator2 
\citep{euclid2021euclid}, which operate on a single resolution of simulations. 
\cite{ho2022multifidelity,ho2023mfbox} use low-fidelity runs across cosmologies 
to approximate high-fidelity simulation output at new cosmological settings. 
Our approach differs fundamentally in scientific objective: rather than predicting 
what a high-resolution run would produce at a new cosmology, we synthesize 
perturbation theory, low-resolution, and high-resolution runs within each cosmology 
to estimate the physically-motivated infinite volume spectrum $P_\infty(k)$, 
trusting each data source only over the wavenumber range where it is approximately unbiased.

For each cosmology, Mira-Titan contains a batch of 18 simulated power spectra: an inexpensive 
power spectrum estimated from perturbation theory ($y_p$), 16 power spectra estimated from 
low resolution simulations ($y_{\ell_r}, r \in \{1,\dots,16\}$), and one power spectrum from 
a high resolution simulation ($y_h$). 
These power spectra are represented as a function of wavenumber $k$, expressed in units of 
$\text{Mpc}^{-1}$.  Following 
\cite{moran2023mira}, we work with a transformation of the original spectra,
$\mathcal{P}(k)=\log_{10}\left(\frac{k^{1.5}P(k)}{2\pi^2}\right)$, where $P(k)$ represents the
original spectra (see  Figure \ref{fig:plot_raw_vs_scrP}). This transformation accentuates 
baryonic acoustic oscillations (periodic fluctuations in matter density imprinted by 
pressure waves in the early universe) and stabilizes the modeling. Output is available 
for $n=351$ values of $k$, spanning $0.001 \leq k \leq 5 \text{ Mpc}^{-1}$, 
but each data type has unique wavenumber ranges where it is deemed approximately unbiased. 
For $k<0.04$, only $y_p$ provides a reliable estimate of the infinite resolution spectra. 
$y_{\ell_r}$ are valid for $0.04 \leq k \leq 0.25$, and $y_h$ is valid for $0.04 \leq k \leq 5$.  
Figure \ref{fig:plot_raw_vs_scrP} shows an example of the output for a particular cosmology. 
When we focus on $k$ values where $y_{\ell_r}$ is valid (Figure \ref{fig:plot_data}), 
the correlated fluctuations of the low-resolution and high-resolution runs across 
wavenumber values becomes apparent.

Notice in both panels of Figure \ref{fig:plot_data} that the curves 
become progressively tighter with increasing $k$, reflecting higher precision in both 
the low- and high-resolution spectra. We would like to conveniently wrap this domain-specific 
knowledge inside our Bayesian framework later.  To this end, \cite{moran2023mira} 
used the multiple low- and high-resolution spectra across all cosmologies to
obtain precision estimates, $p_1,\dots, p_n$, across the $n$ 
wavenumber values using a log-log regression model. A multiplier $c\approx 3.73$ for 
the increase in precision from the low- to high-resolution output is also 
estimated with this regression model. 
We use these estimates to define three 
diagonal precision matrices, one for each data type (perturbation theory, 
low resolution average, and high resolution). Specifically, denote $\Lambda_p$, 
$\Lambda_\ell$, and $\Lambda_h$ as the $n\times n$ diagonal matrices with elements
\begin{equation}\label{eq:lambda}
\Lambda_p^{(ii)} = \begin{cases}
    10^8 &\text{for } k_i < 0.04 \\
    0  &\text{otherwise}\\
    \end{cases}
\quad
\Lambda_\ell^{(ii)} = \begin{cases}
    16p_i &\text{for } 0.04 \leq k_i < 0.25 \\
    0  &\text{otherwise}\\
    \end{cases}
\quad
\Lambda_h^{(ii)} = \begin{cases}
    cp_i &\text{for } 0.04 \leq k_i < 5 \\
    0  &\text{otherwise}\\
    \end{cases}
\end{equation}
for $i=1,\dots, n$.  In $\Lambda_p$, we use the sufficiently high precision of $10^8$ 
to ensure the estimated infinite volume spectrum closely matches the linear theory 
spectrum at low values of $k$ (i.e., it enforces agreement to approximately three 
decimal points). In $\Lambda_\ell$, the multiplication by 16 accounts for the sample 
size of the low resolution runs.
We then calculate a weighted average for each cosmology: 
$\bar y = \Lambda^{-1}(\Lambda_p y_p + \Lambda_{\ell} \bar{y}_\ell + \Lambda_h y_h)$, 
where $\Lambda = \Lambda_p + \Lambda_\ell + \Lambda_h$ and 
$\bar{y}_\ell = \frac{1}{16}\sum_{r=1}^{16} y_{\ell_r}$. In Figure \ref{fig:plot_data},
the weighted average is shown in dashed orange.  As an estimate of the
underlying infinite resolution spectrum, the weighted average fluctuates too drastically.  
This is especially apparent after subtracting a LOESS-smoothed weighted average (right panel).
Although the weighted average incorporates all available observations and 
expert knowledge regarding the precisions, it fails to account for the spatial 
dependence inherent in the smooth but stochastic spectra, which is pivotal to 
effectively inferring the smooth spectrum. The diagonal matrices $\Lambda_p$, 
$\Lambda_\ell$, and $\Lambda_h$ encode pointwise precision at each 
wavenumber but do not capture dependence across wavenumbers; this is 
addressed in the Bayesian hierarchical model of Section \ref{sec:hm_fit}.

\section{Bayesian Hierarchical Modeling for Particular Cosmologies}
\label{sec:hm_fit}

Here we describe our main contribution: the use of a Bayesian hierarchical model 
to estimate the underlying spectrum and quantify uncertainty for a particular cosmology.
Let input $X$ represent the vector with elements $x_i = \log_{10}(k_i)$ for $i=1,\dots,n$.
In the previous section, we used lowercase $\bar{y}$ to denote 
the observed weighted average; now, we use $\bar{Y}$ to indicate the corresponding 
random variable within the statistical model.  
In practice, the observed realization $\bar{y}$ is standardized by subtracting its 
mean and dividing by its standard deviation; precision matrices are adjusted 
accordingly to reflect the scaled responses.

\subsection{Model Specification}
\label{sec:mod_spec}

Let $S$ represent the underlying infinite resolution matter power spectrum for a 
particular cosmology.  We consider $\bar{Y}$ a noisy realization of $S$.
Specifically, we assume $\bar{Y}$ is a random realization of a Gaussian 
process centered at $S$ with some covariance, i.e., 
$\bar{Y}\mid S \sim \mathrm{GP}(S, \Sigma_\varepsilon)$.  While previous works
have considered diagonal $\Sigma_\varepsilon$ \citep{moran2023mira}, we find it
essential to account for the smoothly varying spatial dependence across wavenumber values 
(as visualized in Figure \ref{fig:plot_data}).  We thus specify a dense 
covariance matrix $\Sigma_\varepsilon$ that simultaneously
allows us to model the correlated errors and build domain-matter expertise
into the prior. In general, $\Sigma_\varepsilon$
takes the form of a Mat\'ern kernel \citep{stein1999interpolation} whose parameters 
are informed by precision estimates derived from the available simulation runs. 
The specific form of $\Sigma_\varepsilon$ varies by dataset and is detailed 
explicitly for the simulation study, CAMB, and Mira-Titan settings in Sections 
\ref{subsec:sim}, \ref{subsec:camb}, and \ref{subsec:mira_fit}, respectively.

We next place a Gaussian process prior on $S$, i.e.,
$S\sim \mathrm{GP}\left(\boldsymbol{\mu}_S, \Sigma_S(\cdot)\right)$, again with a 
Mat\'ern kernel. Here, we set $\boldsymbol{\mu}_S=\mathbf{0}$, although our model
is applicable to any selection of $\boldsymbol{\mu}_S$. Standard practice would 
use locations $X$ as the inputs to kernel $\Sigma_S(\cdot)$, yet we have found that 
prior to lack sufficient flexibility for our motivating
application.  In the Mira-Titan setting, some nonstationarity is expected due to 
baryonic acoustic oscillations when $-1.4 \leq \log_{10}(k) \leq -0.8$. To account 
for this, we incorporate a latent Gaussian process that we force to be monotonic
following \citet[][denoted ``monoGP'']{barnett2025monotonic}.  This latent layer
accommodates nonstationarity by warping $X$ via 
$W \sim \mathrm{monoGP}\left(\boldsymbol{\mu}_W, \Sigma_W(X)\right)$.
$W$ is also an $n$-vector whose entries $w_i$ correspond to warped versions of each $x_i$.
Again, we set $\boldsymbol{\mu}_W = \mathbf{0}$ without loss of generality (we have
found $\boldsymbol{\mu}_W = X$ works similarly).
Ultimately, this yields the following hierarchical model:
\begin{equation}\label{eq:dgphm}
\begin{aligned}
\bar{Y}\mid S,\Sigma_\varepsilon &\sim \mathrm{GP}(S,\Sigma_\varepsilon) \\
S\mid W &\sim \mathrm{GP}\left(\mathbf{0}, \Sigma_S(W)\right) \\
W &\sim \mathrm{monoGP}\left(\mathbf{0}, \Sigma_W(X)\right).
\end{aligned}
\end{equation}
The functional composition of GPs in the latter two lines forms a ``deep Gaussian process'' 
\citep{damianou2013deep,dunlop2018deep}. Additional latent layers could be considered, 
although we find one is sufficient.

For both $\Sigma_S$ and $\Sigma_W$, we use Mat\'ern kernels with smoothness of $2.5$,
unit scale, and unique lengthscales:
\begin{equation}\label{eq:cov}
\begin{array}{l}
\Sigma_S(W)^{i,j} = K\left(||w_i - w_j||^2, \theta_S\right) \\[7pt]
\Sigma_W(X)^{i,j} = K\left(||x_i - x_j||^2, \theta_W\right)
\end{array}
\quad\textrm{where}\quad
K(d, \theta) = \left( 1 + \frac{\sqrt{5}d}{\sqrt{\theta}} + 
  \frac{5d^2}{3\theta}\right) \exp\left(-\frac{\sqrt{5}d}{\sqrt{\theta}}\right).
\end{equation}
In our one-dimensional setting, we embrace unit scale on $S$ to preserve parsimony 
and identifiability with $\theta_S$ \citep{zhang2004inconsistent}.
Unit scale on latent $W$ was recommended by \citet{sauer2023active} for similar reasons.

\subsection{Bayesian Inference}\label{sec:inference}

Our inferential goal is to obtain the posterior distribution of 
$S\mid\bar{Y},\Sigma_\varepsilon$, but this requires inferring latent $W$.  
We first integrate over $S$ to condense our hierarchical model of Eq.~(\ref{eq:dgphm}) into
\begin{align}
\label{eq:likelihood}
\bar{Y} \mid W, \Sigma_\varepsilon &\sim \textrm{GP}(\mathbf{0}, \Sigma_S(W) + \Sigma_\varepsilon) \\
\label{eq:prior}
W &\sim \mathrm{monoGP}\left(\mathbf{0}, \Sigma_W(X)\right).
\end{align}
A detailed derivation, including generalizations for 
$\boldsymbol{\mu}_S\neq\mathbf{0}$, is provided in \ref{sec:apdx_int_lik}. 
The incorporation of $\Sigma_\varepsilon$ in the
Gaussian likelihood of this outer layer is an essential upgrade to previous DGP inferential schemes.
This covariance incorporates knowledge of the smoothly varying dynamics from which $\bar{Y}$ was
generated.

We embrace fully-Bayesian inference of latent $W$ and kernel hyperparameters $\{\theta_S, \theta_W\}$.
Specifically, we use elliptical slice sampling \citep[ESS;][]{murray2010elliptical} to 
generate posterior samples of $W$, using Eq.~(\ref{eq:prior}) for proposal generation and 
Eq.~(\ref{eq:likelihood}) for likelihood-based acceptance.  
For $\theta_S$ and $\theta_W$, after pre-scaling, we adopt the default prior specifications 
provided in the {\tt deepgp} {\sf R}-package \citep{deepgp}.
We then integrate Metropolis-Hastings sampling of these parameters in a Gibbs (i.e., one parameter 
at a time updating) framework with the ESS sampling of $W$ \citep{sauer2023active}.

Given burned-in posterior samples $W^{(t)}$ and $\theta_S^{(t)}$
for $t\in\mathcal{T}$ ($\theta_W^{(t)}$ is only used to sample $W^{(t)}$), 
we may leverage Bayes' Theorem to obtain the
posterior distribution of the infinite resolution matter power spectrum for a particular cosmology as
\[
S^{(t)}\mid\bar{Y}, \Sigma_\varepsilon \sim \mathrm{GP}(m^{(t)}, C^{(t)})
\quad\textrm{where}\quad
\begin{array}{rl}
C^{(t)}&=\left(\Sigma_S(W^{(t)})^{-1}+\Sigma_\varepsilon^{-1}\right)^{-1} \\
m^{(t)}&=C^{(t)}\left(\Sigma_\varepsilon^{-1}\bar{Y}\right).
\end{array}
\] 
This form is simplified for $\boldsymbol{\mu}_S=\mathbf{0}$; a general derivation 
is provided in \ref{sec:apdx_SgivenY}.  This closed-form enables direct posterior
sampling of $S^{(t)}$ at discrete locations $x_i$ for $i=1,\dots, n$.  
We draw samples accordingly and compute their average 
to obtain our posterior mean for a particular cosmology. 
Additionally, we quantify uncertainty by finding the 2.5th and 97.5th percentiles 
of the posterior draws at each index to obtain a 95\% credible interval.
We denote our inference procedure for the hierarchical model of Eq.~(\ref{eq:dgphm}) as
``DGP.FCO'' given its deep Gaussian process foundation with additional hierarchical
structure to incorporate functional correlated outputs (FCO). 
The sampling scheme is provided as an algorithm in \ref{sec:apdx_algo}.

\subsection{Simulation Study}
\label{subsec:sim}

We compare DGP.FCO to competing models with a simulation study where the underlying 
true curve is known. Following \cite{moran2024dpc}, we consider two test functions 
defined as
\[
\begin{aligned}
  f_1(x) &= m_1 \exp\left(-\frac{u_1x}{2}\right) * 
  \cos\left(x\sqrt{25-\left(\frac{u_1}{2}\right)^2}\right) - \frac{m_1x}{5}\\
  f_2(x) &= \exp\left(-m_2(x-3)^2\right) + 
  \exp\left(-u_2(x-1)^2\right)-0.05\sin\left(8(x-1.9)\right).
\end{aligned}
\]
For additional variation, the values of $m_1, m_2, u_1,$ and $u_2$ are randomly selected from
$m_1 \sim \text{Uniform}(0.5,1.5)$, $u_1 \sim \text{Uniform}(1.5,2.5)$, and $m_2,u_2 
\sim \text{Uniform}(0.6,1.4)$. The outputs of $f_1$ and $f_2$ (conditional on $m_1,u_1$ and 
$m_2,u_2$), determine the true function values for each simulation.
      
\begin{figure}
    \centering
    \includegraphics[width=\textwidth]{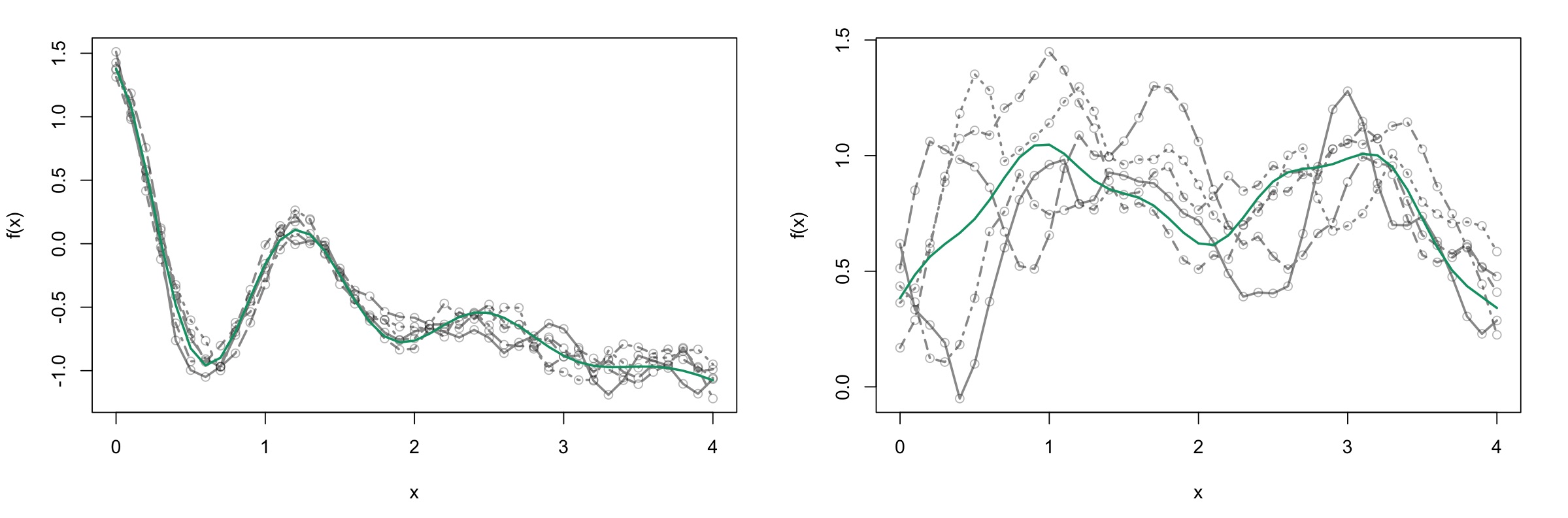}
    \caption{Illustration of simulation output, each with $r=5$ function realizations. The true function
             is shown in solid green, while the gray lines and points indicate the realizations. 
             {\it Left:} $f_1$ with variance setting $A$. 
             {\it Right:} $f_2$ with variance setting $B$.}   
    \label{fig:run_sim_f1A_f2B}
\end{figure}

For each of the two true functions, we consider two variance settings $A$ and $B$, where 
matrices $\Sigma_A$ and $\Sigma_B$ have a Mat\'ern kernel with smoothness of 2.5,
as defined in Eq.~(\ref{eq:cov}). 
We obtain function realizations at $x=\{0, 0.1, \dots, 
3.9, 4.0\}$, so that each function draw contains $n=41$ observations. We draw $r$-many 
realizations from a GP with mean $f_1$ or $f_2$ and spatially varying covariance
$\Sigma_A$ or $\Sigma_B$, where:  

\[
\begin{aligned}
\Sigma_A &= 0.0225 * K(||x_i - x_j||^2, 0.01), \\
\Sigma_B &= s^\top Ms \quad\textrm{where}\quad M = 0.1 * K(||x_i - x_j||^2, 0.05)
\quad\textrm{and}\quad s = 1.5^{-x/2}.
\end{aligned}
\]
Observations are kept noise-free, but we do add a jitter of $10^{-8}$ to the diagonal of $\Sigma_A$ 
and $\Sigma_B$ for numerical stability.  Within $\Sigma_B$, $s$ introduces a degree of
nonstationarity, more effectively mimicking the behavior of the 
CAMB and Mira-Titan datasets.  Examples of simulation output are shown in 
Figure \ref{fig:run_sim_f1A_f2B}.

\begin{figure}[t]
    \centering
    \includegraphics[width=\textwidth]{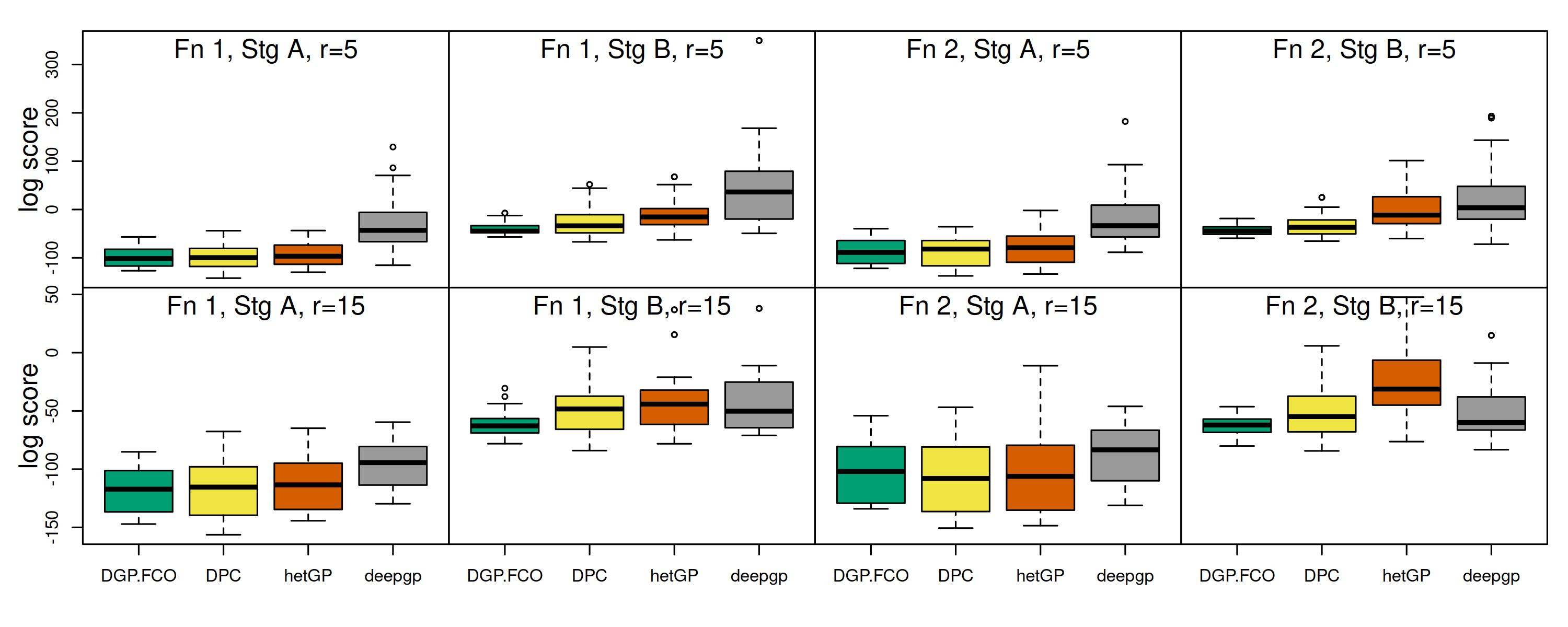}
    \caption{Boxplot of log scores (20 repetitions, lower is better) for each 
             simulated scenario.  Each column represents a function/covariance 
             pair, and each row shows results for $r=5$ (top) or $r=15$ (bottom).}
    \label{fig:sims_logS}
\end{figure}

For each simulated function and variance specification, we conduct 20 Monte Carlo repetitions
with re-randomized $m_1, u_1, m_2, u_2$.  We consider $r = 5$ and $r = 15$.
We compare to an out-of-the-box DGP from the 
\texttt{deepgp} package \citep{sauer2023active}, a heteroskedastic GP from
the \texttt{hetGP} package \citep{binois2018practical, binois2021hetgp}, and
a deep process convolution (DPC) approach, which is utilized within the Cosmic 
Emu \citep{moran2023mira}.  These competing methods view the training data as 
noisy observations with no spatial dependence (i.e., they view the gray dots of
Figure \ref{fig:run_sim_f1A_f2B} without the lines). In
contrast, our method leverages the fact that the observations represent $r$-many
smooth functional realizations in order to estimate parameters
within $\Sigma_\varepsilon$ that account for the spatial dependence.
Our method achieves this by first pre-scaling the functional realizations by the
precision values (to account for nonstationarity), and then using maximum likelihood
estimation to obtain estimates of the marginal variance and lengthscale of the 
Mat\'ern spatial correlation function. We use these estimates and the precision 
information to construct $\Sigma_\varepsilon$, a dense Matérn covariance matrix 
reflecting the smoothly varying spatial dependence across the functional realizations.

We evaluate performance of each method using log score, a proper scoring 
rule which takes both accuracy and UQ into 
consideration \citep[][lower is better]{gneiting2007strictly}.
Results are shown in Figure \ref{fig:sims_logS}.  Our DGP.FCO 
model performs favorably across the board; accounting for the spatial 
dependence between the observations results in more effective UQ.
We relegate mean squared error (MSE) results to \ref{sec:apdx_sims}, because
MSE neglects UQ and was comparable across competing methods.

\subsection{CAMB Estimation}
\label{subsec:camb}

\begin{figure}
    \centering
    \includegraphics[width=\textwidth]{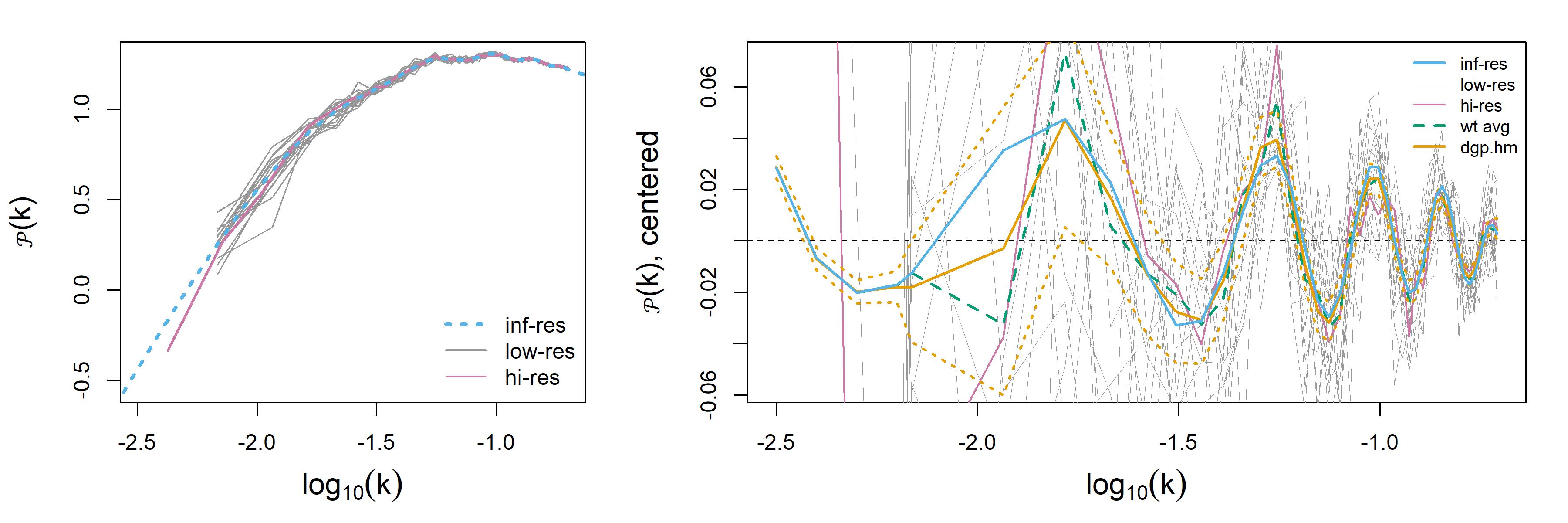}
    \caption{{\it Left:} The infinite-resolution (dotted blue), 
             low-resolution runs (solid gray), and high-resolution run (solid pink) 
             for the first CAMB cosmology.
             {\it Right:} Example of DGP.FCO model fit for the first CAMB cosmology, 
             with a LOESS mean of the 
             weighted average subtracted to emphasize details. We can 
             compare the orange lines (posterior mean and dotted 95\% credible interval) to the 
             light blue line, representing the infinite-resolution ``true" spectrum.}   
    \label{fig:fit_camb}
\end{figure}

The CAMB \citep{lewis2011CAMB} dataset contains matter power spectra based
on six cosmological parameter settings. It includes sixty-four cosmologies,
each containing fifteen low-resolution spectra ($y_{\ell_r}$) and one high-resolution spectrum ($y_h$).
Each spectrum is observed along the same grid of wavenumbers ($k_i$ for $i=1,\dots,n$). 
Unlike the Mira-Titan suite, this dataset includes an infinite-resolution 
spectrum ($y_c)$ which can be treated as the true spectrum for each cosmology,
allowing us to assess model performance. An illustration of the first cosmology 
from CAMB is shown in Figure \ref{fig:fit_camb}.  The ``true'' infinite-resolution 
spectrum is notably smoother than its low- and high-resolution counterparts.

To imitate the role of the perturbation theory output of Mira-Titan, we anchor 
estimates of the CAMB power spectrum at the infinite-resolution for 
$\log_{10}(k) < -2.2$, and then rely strictly on the low- and high-resolution 
runs for $k$ above this (note the transition
in Figure \ref{fig:fit_camb} at $\log_{10}(k) = -2.2$). To compute $\bar{y}$ and $\Sigma_\varepsilon$,
we first use a log-log regression model to obtain precision estimates $\Lambda_\ell$ and $\Lambda_h$
for the low- and high-resolution runs respectively, as outlined in Section \ref{sec:data}. 
To anchor estimates at low resolutions, we define the diagonal matrix $\Lambda_c$ with 
$\Lambda_c^{(ii)}=10^8$ for $\log_{10}(k_i) < -2.2$ ($0$ otherwise).
We then compute the weighted average as: $\bar y = \Lambda^{-1}(\Lambda_c y_c + 
\Lambda_{\ell} \bar{y}_\ell + \Lambda_h y_h)$, where $\Lambda = \Lambda_c + \Lambda_\ell + \Lambda_h$ 
and $\bar{y}_\ell = \frac{1}{15}\sum_{r=1}^{15} y_{\ell_r}$.
For the CAMB data, spatial correlation decays very quickly
(compare the smoothly varying low-resolution runs of Mira-Titan in Figure \ref{fig:plot_data} 
with the jagged low-resolution runs of CAMB in Figure \ref{fig:fit_camb}). 
Although we considered a Mat\'ern kernel for $\Sigma_\varepsilon$ here, we found the covariance
to decay so quickly ($\theta\approx 0$) that a diagonal covariance would suffice.
Therefore, we drop spatial dependence and specify $\Sigma_\varepsilon = 
(\Lambda_c+\Lambda_\ell+\Lambda_h)^{-1}$.  

We fit our DGP.FCO model to the low- and high-resolution runs of each CAMB cosmology for
later use in Section \ref{sec:pred}.  For now, we provide an illustration of the DGP.FCO
prediction (solid yellow) and credible intervals (dashed yellow) for a single 
CAMB cosmology in Figure \ref{fig:fit_camb}.  Our posterior mean provides an effective estimate
of the ``true'' infinite-resolution run.  It aptly smooths over the peaks where 
both the high-resolution run and the weighted average significantly deviate from the truth
($\log_{10}(k)\approx -1.8,-1.3$). Our model also provides effective UQ, appropriately shrinking
credible intervals as precision increases for larger $k$.

\begin{figure}[!b]
    \centering
    \includegraphics[width=.45\textwidth]{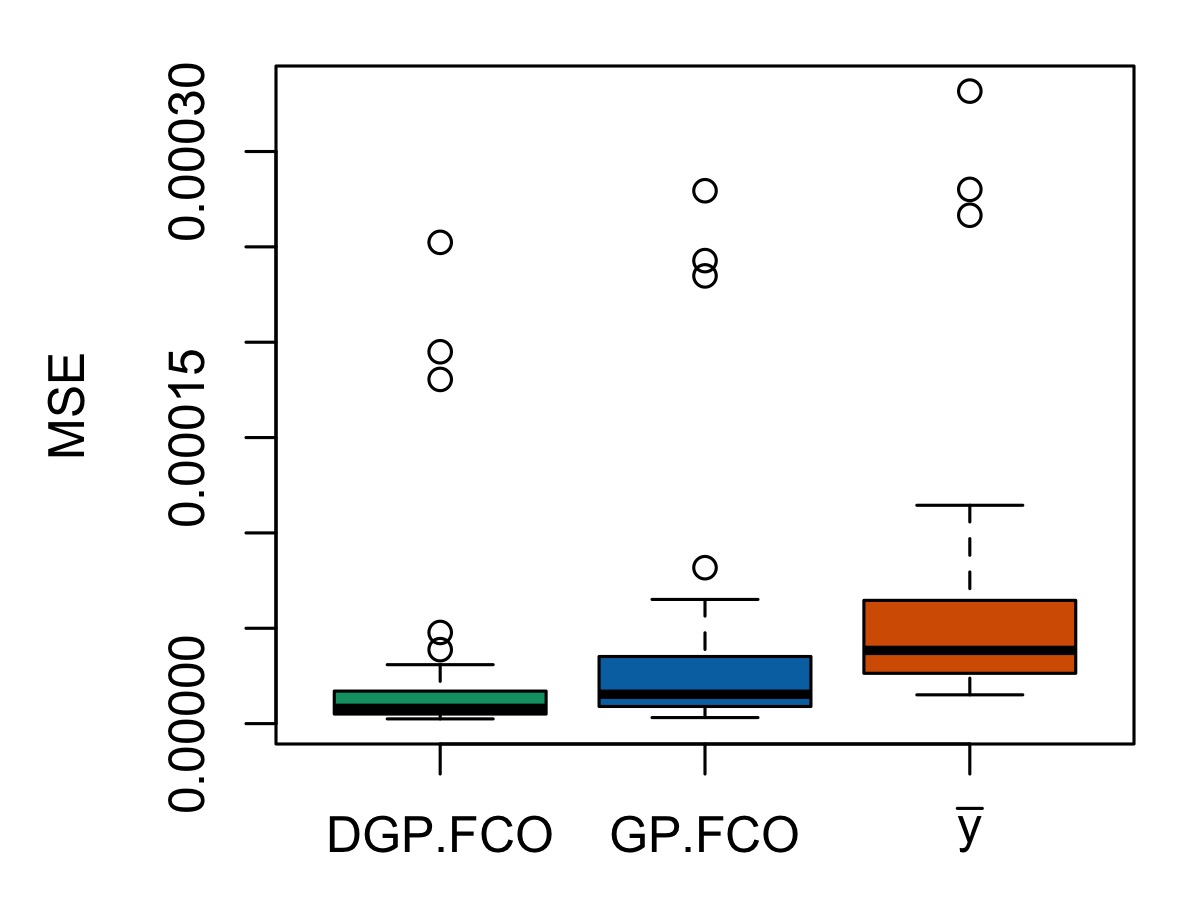}
    \caption{Boxplot of average MSEs (lower is better) across 32 CAMB cosmologies for 
    three estimation methods: the posterior mean from DGP.FCO, the posterior mean from 
    a standard GP leveraging correlated functional outputs (GP.FCO), and the weighted 
    average $\bar{y}$. These estimates are compared against the infinite-resolution 
    spectrum $y_c$ for each cosmology.}
    \label{fig:CAMB_MSEs_box}
\end{figure}

To further validate the necessity of both the smoothing step and 
the deep GP layer, we compare three approaches against the infinite-resolution 
spectrum $y_c$ across all 32 CAMB training cosmologies. Using the weighted average 
$\bar{Y}$ directly yields a mean MSE of $6.728\times10^{-5}$. Replacing $\bar{Y}$ with the 
posterior mean from a standard GP leveraging correlated functional outputs reduces this to 
$4.288\times10^{-5}$, demonstrating the value of the smoothing step. Replacing the standard GP 
with our DGP.FCO posterior mean further reduces the mean MSE to $2.991\times10^{-5}$, 
an additional 30\% reduction, with DGP.FCO outperforming the standard GP on all 32 held-out 
cosmologies. Boxplots of the MSEs (averaged across $k$ values) for each method across the 
32 cosmologies are shown in Figure \ref{fig:CAMB_MSEs_box}. Together, these results 
confirm the value of both the smoothing step and the deep GP layer in recovering the 
infinite volume spectrum. These comparisons serve as validation of the DGP.FCO estimation 
approach, since the infinite-resolution spectrum $y_c$ provides a ground truth unavailable 
in the Mira-Titan setting.

\subsection{Mira-Titan Estimation}
\label{subsec:mira_fit}

Here, we revisit the Mira-Titan suite described in Section \ref{sec:data} to provide
final details for our estimation of the underlying infinite resolution matter power spectrum for
each cosmology.  Recall, with $\Lambda_p$, $\Lambda_\ell$, and $\Lambda_h$ as 
defined in Eq.~(\ref{eq:lambda}), we compute
$\bar y = \Lambda^{-1}(\Lambda_p y_p + \Lambda_{\ell} \bar{y}_\ell + \Lambda_h y_h)$, 
where $\Lambda = \Lambda_p + \Lambda_\ell + \Lambda_h$. Since the low-resolution runs vary 
smoothly across wavenumber values around the underlying spectrum $S$ (as visible in Figure 
\ref{fig:plot_data}), we model their correlation structure using a dense Mat\'ern covariance 
matrix $\Sigma_\ell$ for $0.04 \leq k \leq 0.25$, rather than assuming 
independence across wavenumber values. In previous work, we considered multiple candidates for 
$\Sigma_\varepsilon$ \citep{walsh2023bayesian} and found that a Mat\'ern covariance 
for $\Sigma_\ell$ trained on the low-resolution runs (pre-scaled by the precision 
values and subtracting a LOESS-smoothed average) performs the best. 
We specify this structure within $\Sigma_\varepsilon$ and refer readers 
to \cite{walsh2023bayesian} for a more detailed exposition. This yields 
$\Sigma_\varepsilon=\left(\Lambda_p + \Sigma_\ell^{-1} + \Lambda_h\right)^{-1}$.

\begin{figure}[ht]
    \centering
    \includegraphics[width=\textwidth]{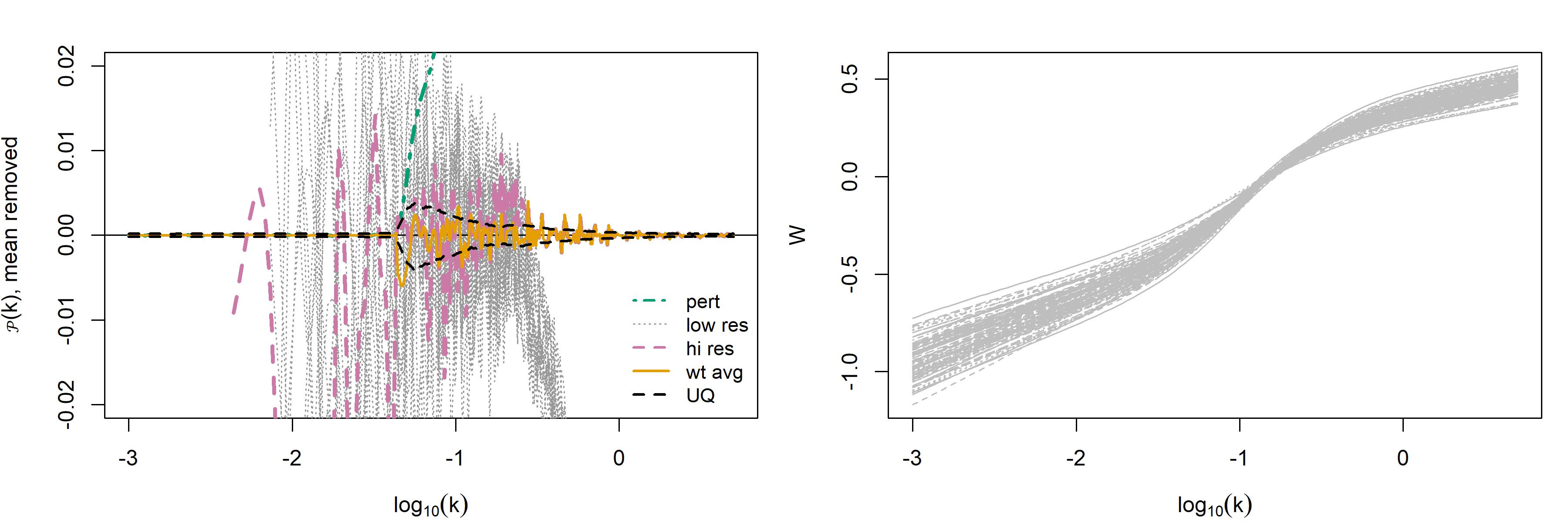}
    \caption{{\it Left:} 95\% credible intervals for the power spectrum of the first Mira-Titan
    cosmology (centered by the predicted posterior mean). Perturbation theory, 
    low-resolution, and high-resolution runs are shown for context. {\it Right:} Posterior draws
    of $W$, illustrating nonstationary behavior by departing from a linear (identity) mapping.}
    \label{fig:plot_fit}
\end{figure}

We then fit our DGP.FCO model on each of 111 training cosmologies, reserving 6 cosmologies
for later hold-out testing.  The left panel of Figure \ref{fig:plot_fit} 
visualizes the DGP.FCO fit for the first training cosmology (compare this with the right
panel of Figure \ref{fig:plot_data}).  We center the plot
by subtracting the posterior mean in order to more easily visualize the credible
interval bounds.  The UQ from the DGP.FCO model accounts for varying precisions 
across differing $k$ values, as well as the spatial correlation within the 
functional realizations.  To provide insight into the nonstationarity of the power
matter spectrum, the right panel of Figure \ref{fig:plot_fit} shows burned-in elliptical 
slice samples of $W$.
These warpings depart from the identity mapping around $x\approx-1$, 
stretching inputs where dynamics are shifting more rapidly for larger $k$ values.

\section{Prediction for Unobserved Cosmologies}
\label{sec:pred}

Here we assess the Bayesian DGP.FCO model posterior means obtained from training cosmologies 
to predict the matter power spectra for held-out cosmologies. 

\subsection{Functional Prediction Model}
\label{subsec:pca}

Let $\psi \in \mathbb{R}^{p_\psi}$ denote the set of parameters that define a particular cosmology.
Mira-Titan has 8 such parameters as described in Section \ref{sec:data}; CAMB has 6.
Our goal is to effectively predict the matter power spectrum as a function of $\psi$, i.e., $\hat{S}(\psi)$, 
for a held-out cosmology.  We fit our DGP.FCO model (Section \ref{sec:inference}) separately to each of the
$m$ training cosmologies. Let $\hat{S}_j$ for $j \in \{1,\ldots,m\}$ denote
the estimated posterior mean of each training cosmology.  In total, our training data comprises
the $p_\psi$-dimensional input $\psi_j$ and the $n$-dimensional output $\hat{S}_j$ (which is a 
function of $x_i = \log_{10}(k_i)$) for $j \in \{1,\ldots,m\}$.

\begin{figure}
    \centering
    \includegraphics[width=\textwidth]{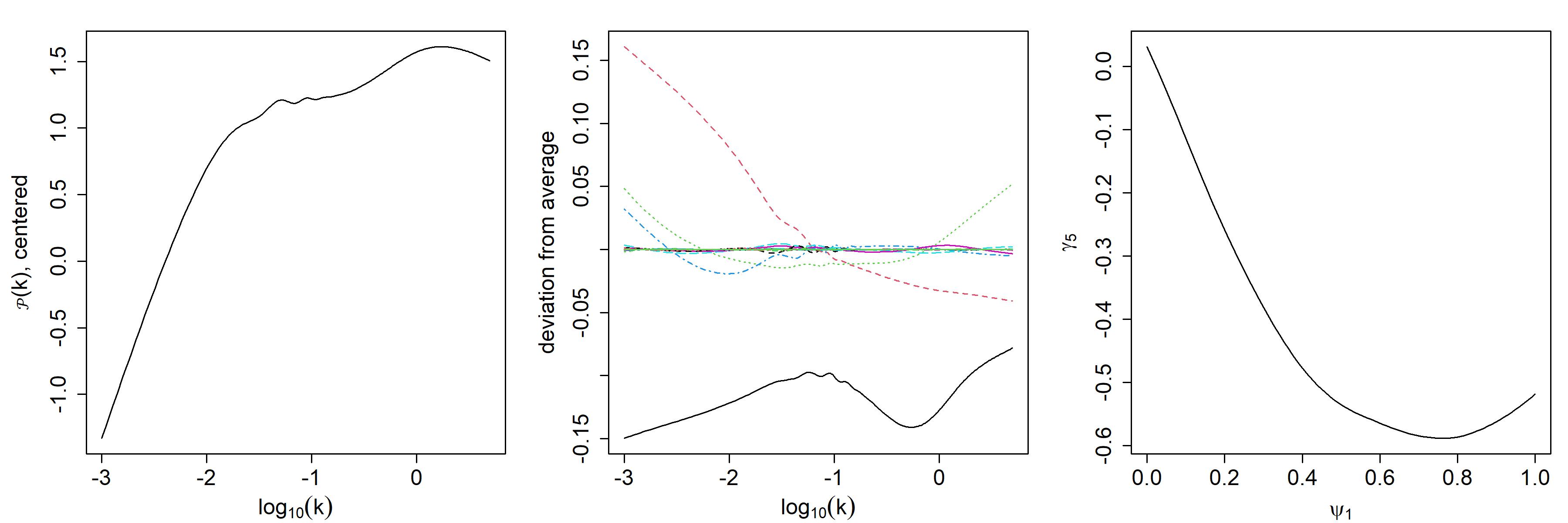}
    \caption{{\it Left:} The estimated mean trend of the $m=111$ posterior means from 
    the Mira-Titan dataset. {\it Middle:} The $p_\eta=10$ principal components obtained from 
    the posterior means. {\it Right:} An 
    illustration of how the weights for the fifth PC ($\gamma_5$) will change as the first 
    cosmological parameter ($\psi_1$) varies.}
    \label{fig:mean_PCs_oneW}
\end{figure}

To handle the functional nature of this output, we reformulate the problem using principal
components \citep[PCs; e.g.,][]{banerjee2014linear}.  
We first center all $\hat{S}_j$ by subtracting $\bar{S} = \frac{1}{m}\sum_{j=1}^{m} \hat{S}_j$, 
then we combine them into an $n\times m$ matrix $\boldsymbol\eta$.  We use the first $p_\eta$ 
principal components to decompose $\boldsymbol\eta = B\Gamma^\top$ where $B$ is the 
$n\times p_\eta$ matrix of basis functions and $\Gamma$ is the $m\times p_\eta$ matrix of basis weights.  
Details on this decomposition are reserved for \ref{sec:apdx_basis}. For both the CAMB and Mira-Titan 
datasets, we set $p_\eta=10$. As a visual, the left panel of Figure \ref{fig:mean_PCs_oneW} shows 
$\bar{S}$ for the Mira-Titan training cosmologies.  The center panel shows the $p_\eta$ basis functions.

Let $\gamma_i^{(j)}$ denote the $ij^\textrm{th}$ element of $\Gamma$, representing the basis weight for the 
$j^\textrm{th}$ cosmology and the $i^\textrm{th}$ PC.
Conditional on $B$, our training data now consists of scalars $\gamma_i^{(j)}$ as a function of $\psi_j$
for $j \in \{1,\ldots,m\}$ and $i \in \{1,\ldots,p_\eta\}$.  If we can effectively predict these basis weights
as a function of $\psi$ (i.e., $\hat{\gamma}_i(\psi)$ for $i \in \{1,\ldots,p_\eta\}$), then we may obtain predictions 
of the power spectrum as
\begin{equation}\label{eq:spred}
\hat{S}(\psi) = \sum_{i=1}^{p_\eta} \hat{\gamma}_i(\psi) B_i,
\end{equation}
where $B_i$ is the $i^\textrm{th}$ basis function (column) of $B$.

In alignment with \cite{higdon2010estcosmo} and \cite{moran2023mira}, we use GPs with a power 
exponential kernel to model the principal components' weights. That is, each vector of PC 
weights $\gamma_i(\psi)$, $i \in \{1,\ldots,p_\eta\}$, is modeled as a zero mean 
GP: $\gamma_i(\psi) \sim \mathrm{GP}(0, \sigma^2R)$, with kernel
\begin{equation*}
    R^{u,v} = \prod_{j=1}^{p_\psi}\exp\left(-10^{\beta_j}|\psi_{uj}-\psi_{vj}|^\alpha\right).
\end{equation*}
We fix $\alpha=1.95$, selected by minimizing MSE across the 
six held-out Mira-Titan cosmologies over a grid of candidate values, and estimate $\sigma^2$ 
and each lengthscale $\beta_1,\ldots,\beta_{p_\psi}$ 
with the \texttt{GPfit} {\sf R}-package \citep{macdonald2015gpfit}, which employs multi-start 
gradient-based maximum likelihood estimation. We repeat this estimation
process for all of the $p_\eta$ PCs.  Conditioned on kernel hyperparameters,
posterior predictions of $\hat{\gamma}_i(\psi^*)$ for a new cosmological parameterization 
$\psi^* \in \mathbb{R}^{p_\psi}$ follows:

\begin{equation*}
\hat{\gamma}_i(\psi^*) = r(\psi^*)^\top R^{-1} \boldsymbol{\gamma}_i,
\end{equation*}

\noindent
where $\boldsymbol{\gamma}_i = [\gamma_i^{(1)}, \gamma_i^{(2)}, \ldots, \gamma_i^{(m)}]^\top$ 
is the vector of $m$ observed PC weights for the $i^\textrm{th}$ component, 
$R$ is the $m \times m$ correlation matrix between 
all pairs of training inputs defined above, and $r(\psi^*) \in \mathbb{R}^m$ is 
the vector of correlations between the test input $\psi^*$ and each training input 
$\psi_j$. To illustrate, the right panel of Figure \ref{fig:mean_PCs_oneW} shows
the predicted weight of the fifth PC ($\gamma_5$) for Mira-Titan as a function 
of the first cosmological parameter ($\psi_1$), while the other parameters are fixed 
at their midpoint.  With all estimated PC weights, $\hat{\gamma_i}(\psi^*), i=1,\ldots,p_\eta$, 
we predict power spectra following Eq.~(\ref{eq:spred}).  

\subsection{Predicting Spectra for CAMB}
\label{subsec:camb_pred}

For the CAMB dataset, we split the 64 available cosmologies such that $m=32$
are used for training and $N=32$ are reserved as hold-out/test cosmologies for prediction. 
For each training cosmology, we collect our DGP.FCO predicted spectrum, then conduct the principal
components model just described to predict $\hat{S}(\psi)$ for each test cosmology.  

\begin{figure}
    \centering
    \includegraphics[width=0.85\textwidth]{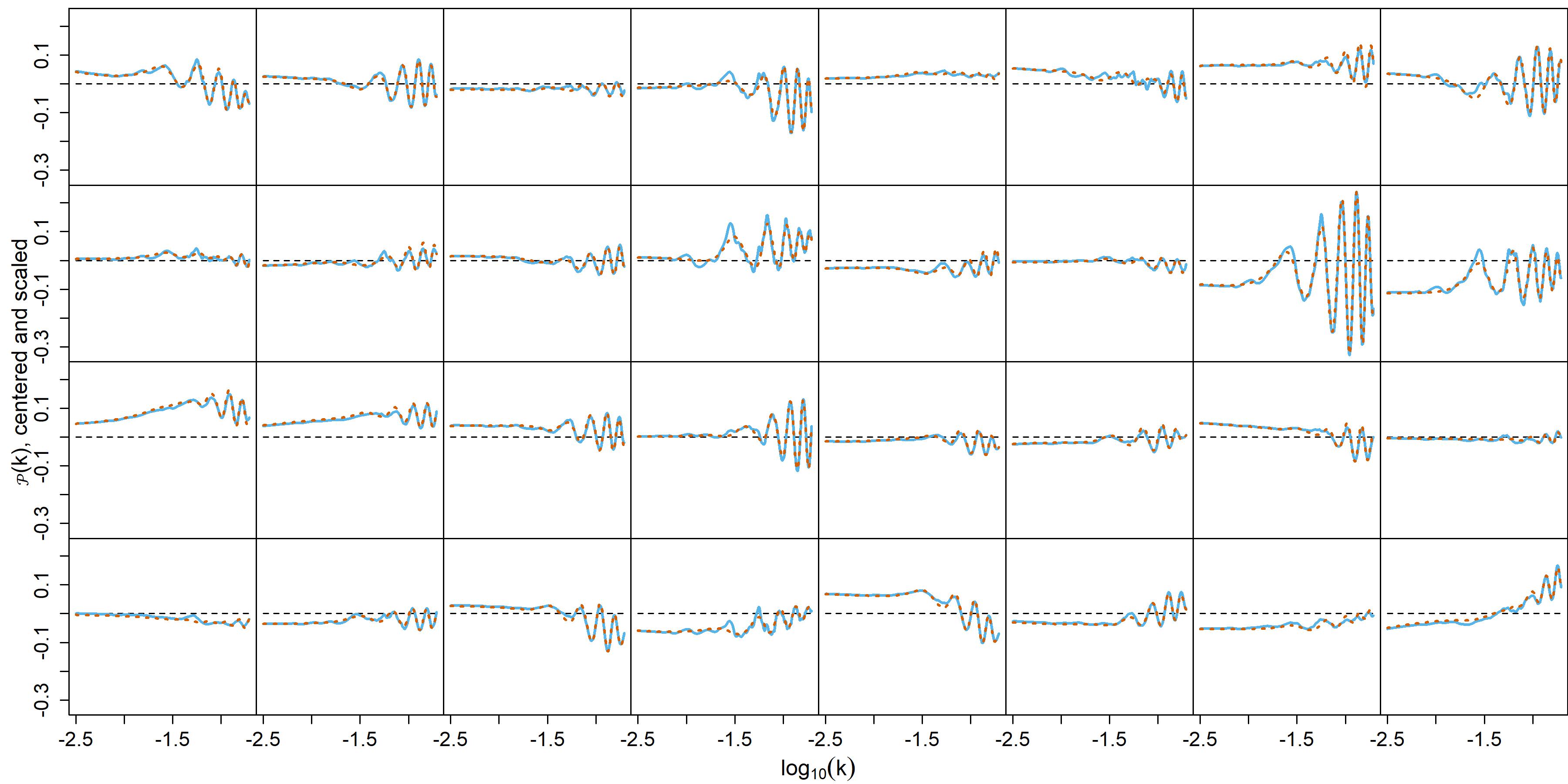}
    \caption{Predictions of matter power spectra for each of the 32 CAMB test cosmologies,
             after centering and scaling (to accentuate differences) and with the true 
             infinite resolution run subtracted (shown as black dashed 0 line), the PC prediction 
             trained on the infinite resolution spectra is in solid blue, and the PC prediction trained
             on the DGP.FCO posterior means is in dotted orange.}
    \label{fig:pca_preds_v_camb}
\end{figure}

\begin{figure}[!t]
    \centering
    \includegraphics[width=2in]{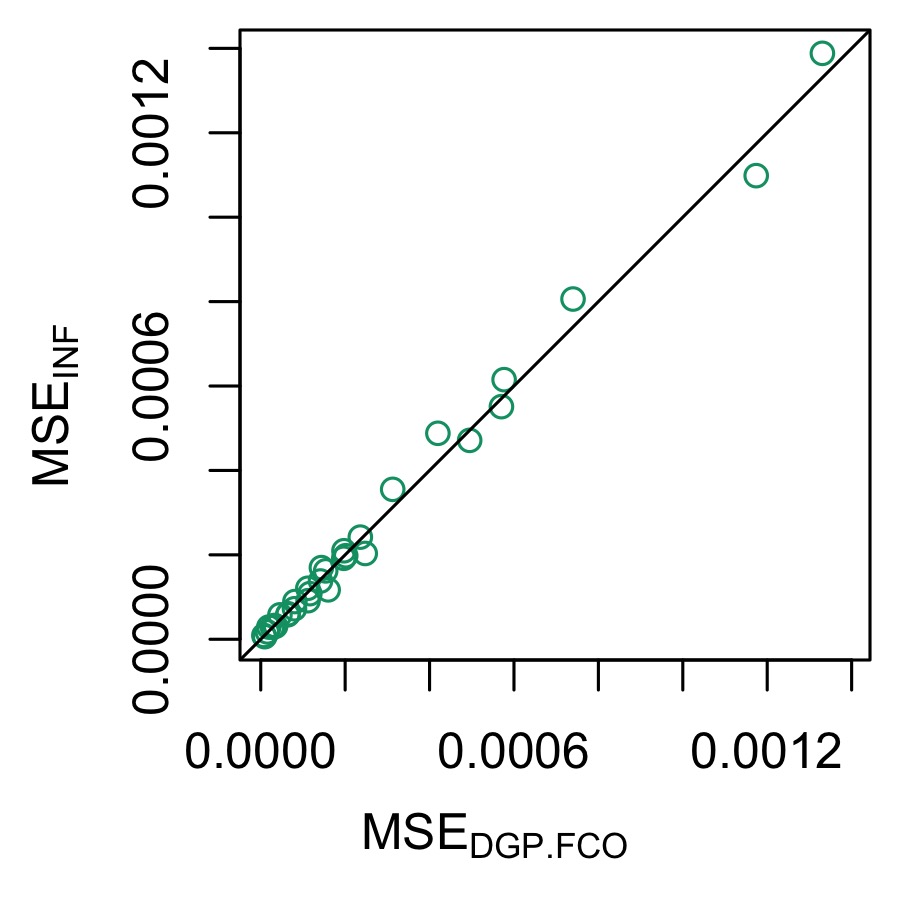}
    \caption{Comparison of MSE for CAMB predictions, when training with DGP.FCO posterior 
             means and training with the ``true" infinite-resolution spectra. The identity line 
             provides equal MSE values among both methods.}   
    \label{fig:mse_camb}
\end{figure}

Since CAMB offers the ``true'' infinite-resolution spectra, we have the opportunity to benchmark our
results in two ways.  First, we may compare our predictions to the true power spectrum for each held-out
cosmology.  Second, we may leverage the infinite-resolution spectra for the training cosmologies in
place of our estimated $\hat{S}_j$ from our DGP.FCO model, to get a sense of the ``best-case" predictions 
that could originate from the PC decomposition.  
Figure \ref{fig:pca_preds_v_camb} compares our predictions (DGP.FCO + PC, dotted orange)
to these ``best-case'' predictions (true + PC, solid blue) with the actual 
infinite-resolution spectra subtracted (dashed black)
for each of the 32 held-out test cosmologies.  
The agreement between our predictions and the ``best-case'' predictions indicates that DGP.FCO is effectively
capturing the true matter power spectra of the training cosmologies.
The high level of agreement between the pairs of predictions and the true spectra across cosmologies
indicates our PC model is providing effective predictions of held-out power spectra.
We then compare the mean squared errors of our predictions and the ``best-case'' 
predictions across all held-out cosmologies, with results shown in Figure \ref{fig:mse_camb}.  We again
see only minimal differences between the two, confirming that DGP.FCO is offering effective
estimates of the true spectra for each cosmology.

\subsection{Predicting Spectra for Mira-Titan}
\label{subsec:mira_pred}

Here we show prediction results for $N=6$ held-out cosmologies for the Mira-Titan dataset. 
We compare our method with Cosmic Emu \citep{moran2023mira}, the state-of-the-art emulator 
constructed on the same $m=111$ Mira-Titan cosmologies. 
Cosmic Emu uses a Bayesian approach 
where the underlying infinite volume spectrum for each cosmology is modeled with a process 
convolution on Brownian motion. 
Nonstationarity is permitted through modeling the bandwidth parameter with a process 
convolution as well; this composition is referred to as a deep process convolution 
\citep[DPC;][]{moran2023mira}. Cosmic Emu also uses a principal components model 
(as described in Section \ref{subsec:pca}) to predict power matter 
spectra for held-out cosmologies as a function of $\psi$ \citep{gattiker2020sepia}.

\begin{figure}[!b]
    \centering
    \includegraphics[width=\textwidth]{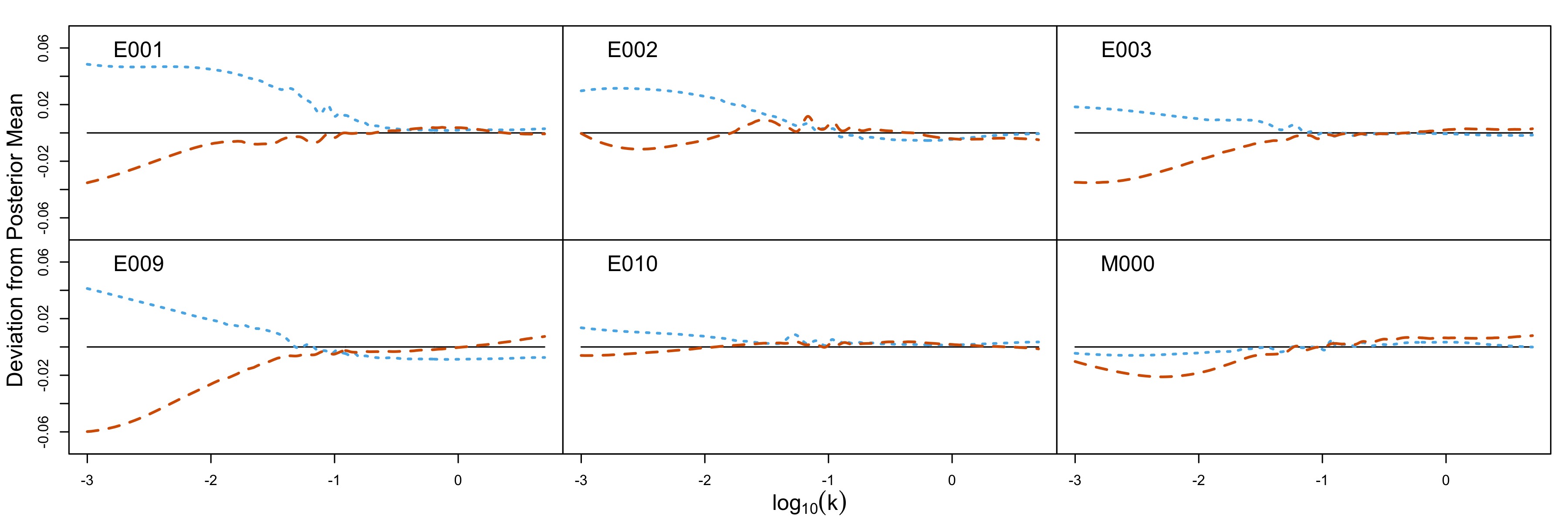}
    \caption{Results of all six predictions for both methods on the held-out 
    cosmologies (respective in-sample posterior mean subtracted). 
    Cosmic Emu is dotted blue and DGP.FCO + PC is dashed orange.
    Closer to the dotted zero line indicates a better fit with the 
    respective posterior mean.}
    \label{fig:plot_pred_1to6}
\end{figure}

Using our proposed DGP.FCO approach and Cosmic Emu, we obtain two sets of predictions 
for $\hat{S}_t(\psi_t)$ with $t\in\{1,\ldots,6\}$ indexing the six held-out cosmologies.  
In this real-world example, we have no ``truth" for the Mira-Titan hold-outs against 
which to benchmark. In an effort to assess the accuracy of these predictions, 
we also separately fit our DGP.FCO model and DPC to smooth each of these testing cosmologies
(leveraging the perturbation theory, low resolution runs, and high resolution run 
to predict the underlying matter power spectrum).  Although these estimated ``in-sample'' 
spectra for each method are not formal ``truths,'' we find them to be
useful benchmarks for our ``out-of-sample'' predictions. The individual predictions 
for each method and held-out cosmology are shown in Figure \ref{fig:plot_pred_1to6}. 
For each method, predictions are centered by their respective in-sample posterior means: 
DGP.FCO + PC predictions are centered by the in-sample DGP.FCO posterior mean, and 
Cosmic Emu predictions are centered by the in-sample DPC posterior mean. 
Predictions closer to zero therefore represent better agreement with each method's 
own estimate of the underlying spectrum.

\begin{figure}
    \centering
    \includegraphics[width=\textwidth]{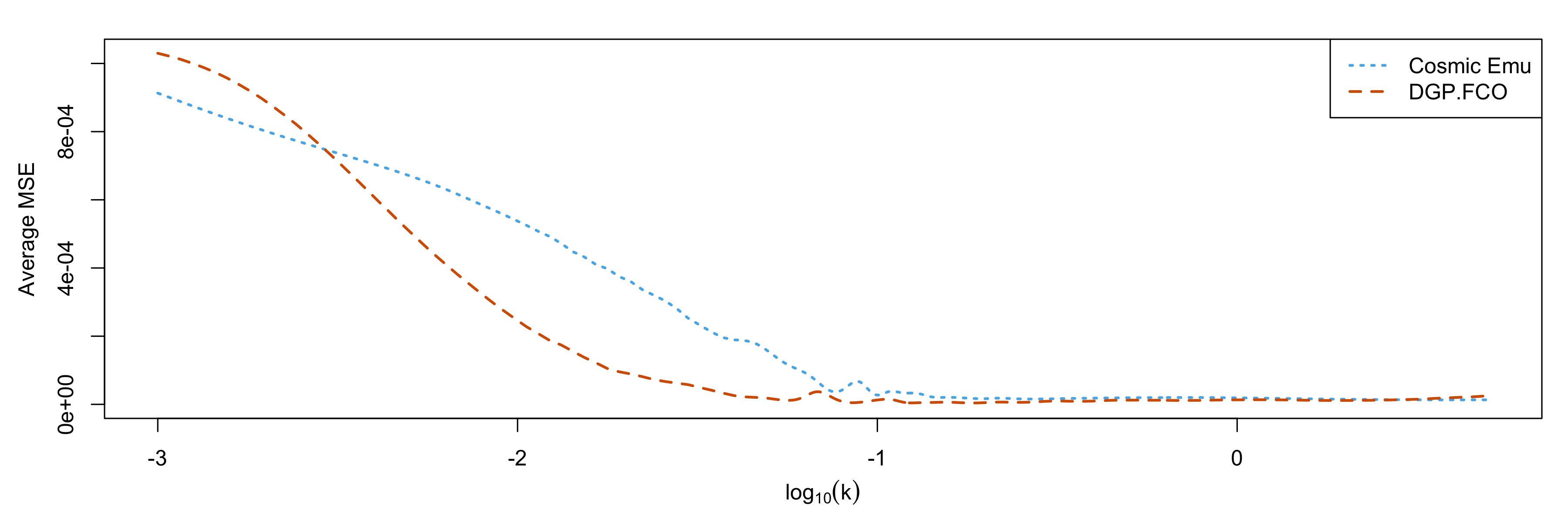}
    \caption{MSE across each of the 6 hold out cosmologies for each $k$ value. Cosmic Emu 
    performance is shown in dotted blue, and DGP.FCO + PC performance is shown in dashed orange.}
    \label{fig:mse_by_k}
\end{figure}

From these predictions, we can calculate MSE against the in-sample posterior means
across the 6 test cosmologies. 
Our out-of-sample DGP.FCO + PC prediction achieves the greatest reduction in MSE
in the region where perturbation theory and the low-resolution runs are deemed unbiased 
(see Figure \ref{fig:mse_by_k}). For the region where only the high-resolution run 
is unbiased, Cosmic Emu has a marginally lower MSE. Our MSE is lower at 88\% of $k$ 
values considered. These results indicate that DGP.FCO competes favorably 
with the state-of-the-art Cosmic Emu model.

\section{Discussion}
\label{sec:disc}

In this work, we introduced a novel Bayesian hierarchical framework that leverages deep 
Gaussian processes (DGPs) to model smooth latent functions from correlated functional outputs, 
such as those arising in cosmological simulations. Our model (DGP.FCO) builds on the 
compositional nature of DGPs and introduces an additional outer layer to explicitly 
model correlated observational errors within each spectrum. Through a simulation study, 
we demonstrated that DGP.FCO offers improved uncertainty 
quantification compared to existing methods.

We applied DGP.FCO to both the CAMB and Mira-Titan datasets, generating posterior mean estimates 
and predicting at held-out cosmological settings. Across both datasets, DGP.FCO 
performs consistently well against competitive baselines, showcasing strong generalization and 
flexibility in complex, functional data regimes.

Several avenues for future research arise. Beyond modeling dark matter power 
spectra, DGP.FCO could be extended to hydrodynamical simulation outputs in cosmology, which 
often share similar functional structures and error correlations. More broadly, this approach may be 
applicable in other scientific domains involving structured functional outputs, especially those 
with spatial or temporal dependencies.

In addition, generating predictions over large cosmological grids opens the door to detailed 
sensitivity analyses, enabling researchers to disentangle parameter interactions and identify 
the most influential physical drivers of the spectra. From a modeling standpoint, further 
developments could include joint modeling of latent layers across cosmologies, and extensions 
to redshifts beyond $z=0$. Estimating hyperparameters for more flexibility of $\Sigma_\varepsilon$ 
within the Bayesian sampling scheme is an area for further investigation. 
Combining the DGP.FCO approach with the multifidelity framework of \cite{ho2023mfbox} 
is another interesting avenue to more comprehensively use simulations of differing 
accuracy and fidelity. Lastly, a hybrid framework combining DGP.FCO with Cosmic Emu, leveraging 
the strengths of each at different scales of $k$, may yield more accurate predictions.

\section{Acknowledgments}
\label{sec:ack}

This work was in part supported by the U.S. Department
of Energy, Office of Science, Office of Advanced Scientific Computing Research, 
Scientific Discovery through Advanced Computing (SciDAC) program through 
Grant 420453 ``Enabling Cosmic Discoveries in the Exascale Era." Work at Argonne 
National Laboratory was supported under the U.S. Department of Energy 
contract DE-AC02-06CH11357.

The authors are pleased to acknowledge Advanced Research Computing at Virginia Tech 
(\url{https://arc.vt.edu/}) as well as the Shared Computing Cluster (SCC) administered
by Boston University's Research Computing Services (\url{www.bu.edu/tech/support/research/})
for providing computational resources and technical support that have contributed 
to the results reported within this paper.
The authors are grateful to the anonymous reviewers for their valuable input and 
for helping to strengthen the manuscript.

\section{Disclosure}
The authors report there are no competing interests to declare. The large language model 
Claude (Sonnet 4.6) was used to assist in rewording and clarifying the manuscript 
during the revision process, as well as debugging R code while reworking images. 
The authors take full responsibility for the accuracy and integrity of the final manuscript.

\section{Data Availability Statement}
We provide data, reproducible code, and an {\sf R} package for our 
Bayesian DGP.FCO model in a public git repository: \url{https://github.com/stevewalsh124/dgp.hm}

\appendix
\renewcommand{\thesection}{Appendix \Alph{section}}

\section{Integrated Likelihood}
\label{sec:apdx_int_lik}
Here, we establish that under the model:
$$
\begin{aligned}
Y_i|S,W &\sim N\big(S, \Sigma_\varepsilon\big), \quad i \in \{1,\dots, r\}\\
S|W &\sim N\big(\boldsymbol{\mu}, \Sigma_S(w,w')\big)\\
W &\sim N\big(\mathbf{0}, \Sigma_W(x,x')\big) \\
\end{aligned}
$$
We can obtain the integrated likelihood $p(Y_i|W)=\int p(Y_i|S,W)p(S|W) dS$, 
with the following calculations:
$$
\begin{aligned}
p(Y_i|W) &= \int p(Y_i|S,W)p(S|W)dS \\
&\propto \int\exp\left(-\frac{1}{2} \left[Y_i^T\Sigma_\varepsilon^{-1} Y_i 
- 2Y_i^T\Sigma_\varepsilon^{-1} S + S^T \Sigma_\varepsilon^{-1} S + S^T \Sigma_S^{-1} S 
-2S^T \Sigma_S^{-1} \boldsymbol{\mu} 
+ \boldsymbol{\mu}^T \Sigma_S^{-1} \boldsymbol{\mu} \right]\right)dS \\
\end{aligned}
$$
If we focus only on terms containing $S$, we have 
$$
\begin{aligned}
    \exp\Big(-\frac{1}{2} \big[S^T &(\Sigma_\varepsilon^{-1}+\Sigma_S^{-1}) S 
    -2S^T (\Sigma_S^{-1} \boldsymbol{\mu} + \Sigma_\varepsilon^{-1} Y_i) \big]\Big) \\
&= \exp\left(-\frac{1}{2} \left[S^T (\Sigma_\varepsilon^{-1}+\Sigma_S^{-1}) S 
-2S^T (\Sigma_\varepsilon^{-1}+\Sigma_S^{-1})(\Sigma_\varepsilon^{-1}
+\Sigma_S^{-1})^{-1}(\Sigma_S^{-1} \boldsymbol{\mu} + \Sigma_\varepsilon^{-1} Y_i) \right]\right)
\end{aligned}
$$
To complete the square, we add and subtract $(\Sigma_S^{-1} \boldsymbol{\mu} 
+ \Sigma_\varepsilon^{-1} Y_i)^T(\Sigma_\varepsilon^{-1}+\Sigma_S^{-1})^{-1}
(\Sigma_S^{-1} \boldsymbol{\mu} + \Sigma_\varepsilon^{-1} Y_i)$, 
which then gives the kernel for $S \sim N\left((\Sigma_\varepsilon^{-1}
+\Sigma_S^{-1})^{-1}(\Sigma_S^{-1} \boldsymbol{\mu} + \Sigma_\varepsilon^{-1} Y_i), 
(\Sigma_\varepsilon^{-1}+\Sigma_S^{-1})^{-1} \right)$. 
This removes the integral and leaves us with the following terms for $Y_i|W$:
$$
\begin{aligned}
&\exp\Big(-\frac{1}{2}\big[Y_i^T\Sigma_\varepsilon^{-1} Y + 
\boldsymbol{\mu}^T\Sigma_S^{-1} \boldsymbol{\mu} - (\Sigma_S^{-1} \boldsymbol{\mu} + 
\Sigma_\varepsilon^{-1} Y_i)^T(\Sigma_\varepsilon^{-1}+\Sigma_S^{-1})^{-1} 
(\Sigma_S^{-1} \boldsymbol{\mu} + \Sigma_\varepsilon^{-1} Y_i)\big]\Big) \\
&= \exp\Big(-\frac{1}{2}\big[Y_i^T\big(\Sigma_\varepsilon^{-1} - 
\Sigma_\varepsilon^{-1}(\Sigma_\varepsilon^{-1}+\Sigma_S^{-1})^{-1}\Sigma_\varepsilon^{-1}\big)Y_i 
-2Y_i^T \Sigma_\varepsilon^{-1}(\Sigma_\varepsilon^{-1}+\Sigma_S^{-1})^{-1}\Sigma_S^{-1} \boldsymbol{\mu} \\
&\qquad \qquad \qquad + \boldsymbol{\mu}^T (\Sigma_S^{-1} - \Sigma_S^{-1}(\Sigma_\varepsilon^{-1}
+\Sigma_S^{-1})^{-1}\Sigma_S^{-1}) \boldsymbol{\mu}\big]\Big) \\
&= \exp\Big(-\frac{1}{2}\big((Y_i-\boldsymbol{\mu})^T(\Sigma_\varepsilon+\Sigma_S)^{-1}
(Y_i-\boldsymbol{\mu}) \big)\Big)
\end{aligned}
$$
With the last equality resulting from the use of the Sherman–Morrison–Woodbury formula, 
yielding $\big(\Sigma_\varepsilon^{-1} - \Sigma_\varepsilon^{-1}(\Sigma_\varepsilon^{-1}+
\Sigma_S^{-1})^{-1}\Sigma_\varepsilon^{-1} = (\Sigma_S^{-1} - \Sigma_S^{-1}
(\Sigma_\varepsilon^{-1}+\Sigma_S^{-1})^{-1}\Sigma_S^{-1}) = (\Sigma_\varepsilon+\Sigma_S)^{-1}$. 
We also use the fact that $\Sigma_\varepsilon^{-1}(\Sigma_\varepsilon^{-1}+
\Sigma_S^{-1})^{-1}\Sigma_S^{-1} = (\Sigma_\varepsilon + \Sigma_S)^{-1}$, 
which holds since it can be shown that 
$[\Sigma_S(\Sigma_\varepsilon^{-1}+\Sigma_S^{-1})\Sigma_\varepsilon](\Sigma_\varepsilon+\Sigma_S)
=(\Sigma_\varepsilon+\Sigma_S)(\Sigma_\varepsilon+\Sigma_S)$.

Therefore, we have that $Y_i|W \sim N\big(\boldsymbol{\mu}, \Sigma_S(w,w')+\Sigma_\varepsilon(w,w')\big)$.

\section{Conditional Distribution of Matter Power Spectrum, Given Weighted Average}
\label{sec:apdx_SgivenY}

From our model specification, we have the following:
\[
S \sim \mathcal{N}(\boldsymbol{\mu}, \Sigma_S)
\]
\[
\bar{Y} \mid S \sim \mathcal{N}(S, \Sigma_\varepsilon)
\]
Then, to find the distribution of $S|\bar{Y}$, we apply Bayes' Theorem below, 
where $\propto$ indicates proportionality:
\[
P(S \mid \bar{Y}) \propto P(\bar{Y} \mid S) P(S)
\]
\[
\propto \exp\left(-\tfrac{1}{2} (\bar{Y} - S)^T \Sigma_\varepsilon^{-1} (\bar{Y} - S) \right) 
\cdot \exp\left(-\tfrac{1}{2} (S - \boldsymbol{\mu})^T \Sigma_S^{-1} (S - \boldsymbol{\mu})\right)
\]
\[
\propto \exp\left(-\tfrac{1}{2} \left[ \bar{Y}^T \Sigma_\varepsilon^{-1} \bar{Y} 
- 2 S^T \Sigma_\varepsilon^{-1} \bar{Y} + S^T \Sigma_\varepsilon^{-1} S + S^T \Sigma_S^{-1} S 
- 2 S^T \Sigma_S^{-1} \boldsymbol{\mu} + \boldsymbol{\mu}^T \Sigma_S^{-1} \boldsymbol{\mu} \right] \right)
\]
\[
\propto \exp\left( -\tfrac{1}{2} \left[ S^T (\Sigma_\varepsilon^{-1} + \Sigma_S^{-1}) S 
- 2 S^T (\Sigma_\varepsilon^{-1} \bar{Y} + \Sigma_S^{-1} \boldsymbol{\mu}) \right] \right)
\]
From properties of the multivariate normal distribution \citep[e.g., Equation 7.1 of][]{hoff2009first},
we can establish that the mean and covariance of $S|\bar{Y}$ will be $\boldsymbol{m}$
and $C$ respectively, where we can solve for these values with:
\[
C^{-1} = \Sigma_\varepsilon^{-1} + \Sigma_S^{-1}
\]
\[
C^{-1} \boldsymbol{m} = \Sigma_\varepsilon^{-1} \bar{Y} + \Sigma_S^{-1} \boldsymbol{\mu}
\]
Then,
\[
C = (\Sigma_\varepsilon^{-1} + \Sigma_S^{-1})^{-1}
\]
\[
\boldsymbol{m} = \left( \Sigma_S^{-1} + \Sigma_\varepsilon^{-1} \right)^{-1} 
\left( \Sigma_\varepsilon^{-1} \bar{Y} + \Sigma_S^{-1} \boldsymbol{\mu} \right)
= C \cdot \left( \Sigma_\varepsilon^{-1} \bar{Y} + \Sigma_S^{-1} \boldsymbol{\mu} \right)
\]
Therefore, we have the result that $S|\bar{Y} \sim \mathcal{N}(\boldsymbol{m}, C)$.

\section{DGP.FCO Bayesian Sampling Scheme Algorithm}
\label{sec:apdx_algo}

Here is the algorithm used to generate posterior samples of $S^{(t)} \mid \bar{Y}, 
\Sigma_\varepsilon$, which further allows calculation of the posterior means for each cosmology 
and corresponding credible intervals.

\hspace{5mm}

\begin{algorithm}[H]
\DontPrintSemicolon
\caption{DGP.FCO Sampling Scheme}

Initialize $W^{(1)}$, $\theta_S^{(1)}$, and $\theta_W^{(1)}$\;

\For{$t = 2, \dots, T$}{
    Sample $W^{(t)}$ via elliptical slice sampling using Eq.~(5) for proposals and Eq.~(4) for likelihood evaluation\;
    
    Sample $\theta_S^{(t)}$ and $\theta_W^{(t)}$ via Metropolis-Hastings within a Gibbs framework\;
}

Discard burn-in samples\;

Given posterior samples $\{W^{(t)}, \theta_S^{(t)}\}$, draw samples of
$S^{(t)} \mid \bar{Y}, \Sigma_\varepsilon$ using the closed-form expression in Section \ref{sec:inference}\;

Compute posterior means by averaging $S^{(t)}$ samples across iterations, and compute pointwise 
$95\%$ credible intervals using the $2.5^{\text{th}}$ and $97.5^{\text{th}}$ posterior percentiles at each index\;

\end{algorithm}

\section{Simulation Study Additional Results}
\label{sec:apdx_sims}

\begin{figure}[H]
    \centering
    \includegraphics[width=\textwidth]{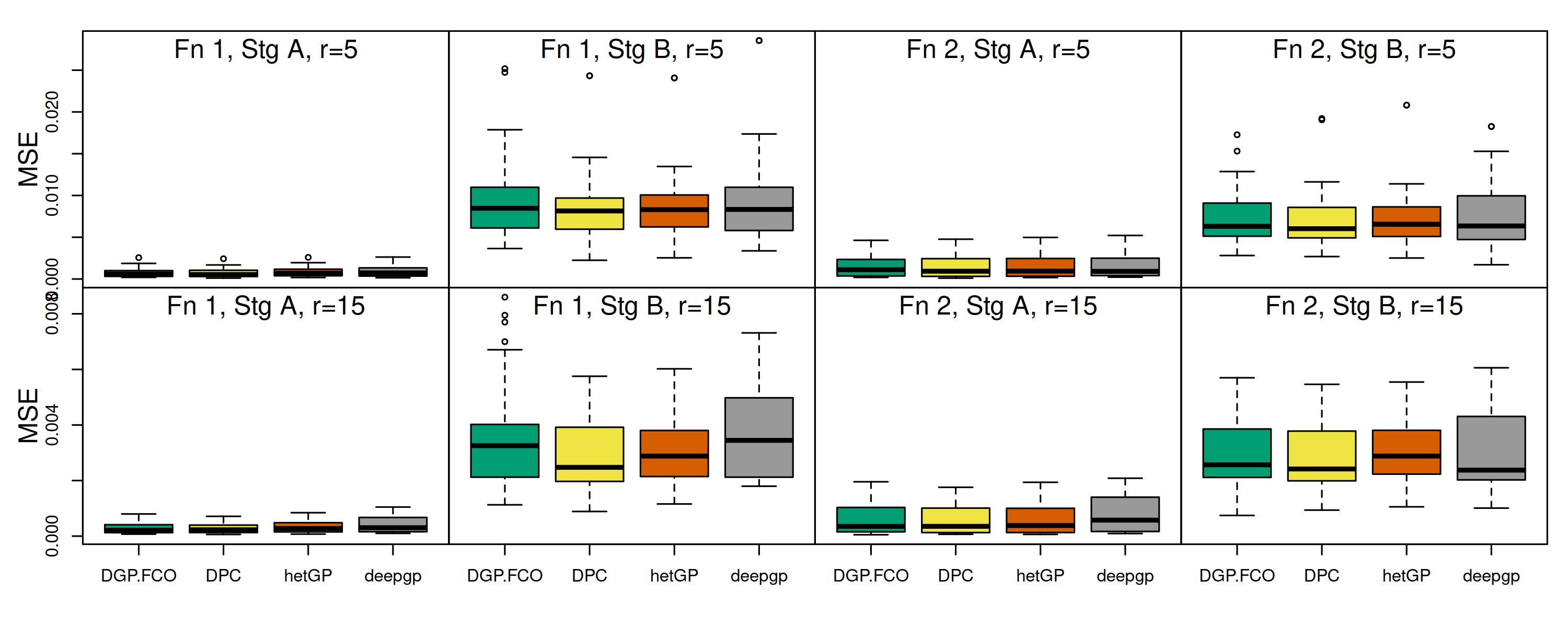}
    \caption{Boxplot of MSEs across two different functions and covariance settings 
             (lower is better). For each case, the boxplot is constructed from 20 
             random batches were simulated. Each column represents a function/variance 
             specification pair, and each row shows results for batch sizes of 
             either 5 (top) or 15 (bottom).}    
    \label{fig:sims_MSE}
\end{figure}

\section{Details on Basis Decomposition for Predictions}
\label{sec:apdx_basis}

Here, we begin with a $n \times m$ matrix composed of $m$ posterior means, 
each of length $n$, denoted by $\boldsymbol\eta$. To facilitate more efficient estimation, 
we use singular value decomposition \citep[SVD; e.g.,][]{banerjee2014linear}, 
where $\boldsymbol\eta = UDV^T$; $U$ and $V$ are orthogonal matrices of size $n \times n$ 
and $m\times m$, respectively. $D$ is a diagonal matrix containing the singular values. 
Here, $\boldsymbol\eta$ has rank $r=\text{min}\{m,n\}$ (for our applications, $r=111$ for Mira-Titan 
and $r=32$ for CAMB), which determines the number of non-zero elements in $D$. We find an alternative 
decomposition where $\Sigma$ is a diagonal matrix with the non-zero singular values of 
$\boldsymbol\eta$, and $U_1$ and $V_1$ are matrices with orthonormal columns of size 
$n_\eta \times r$ and $m \times r$, respectively. 

Following \cite{higdon2008computer, higdon2010estcosmo}, we equivalently decompose 
$\boldsymbol\eta$ into a principal component (PC) basis matrix 
$B^* = \frac{1}{\sqrt{m}}U_1\Sigma$ along with its corresponding weights 
$\Gamma^* = \sqrt{m}V_1$. Without loss of generality, we can perform this decomposition 
after an overall mean trend is removed. If we use $p_\eta < r$ PCs, this will reduce 
$B$ and $\Gamma$ to be the first $p_\eta$ columns of $B^*$ and $\Gamma^*$ respectively and 
result in an approximation where 

\begin{equation}
    \boldsymbol\eta= UDV^T = U_1\Sigma V_1^T \approx B\Gamma^T.
\end{equation}

\bibliographystyle{jasa}
\bibliography{references}

\end{document}